\documentclass[referee,useAMS,usenatbib]{mn2e}
\def\lsim{\lower.5ex\hbox{$\; \buildrel < \over \sim \;$}}
\def\gsim{\lower.5ex\hbox{$\; \buildrel > \over \sim \;$}}
\def\be{\begin{equation}}
\def\ee{\end{equation}}

\def\vel{\vartheta}

\def\md{\dot{\cal M}}

\def\bc{\begin{center}}
\def\ec{\end{center}}
\def\eg{{\it e.g.,}}
\def\etal{{\em et al.}}
\def\ie{{\em i.e.,}}
\def\ep{{{\rm e}^--{\rm p}^+}}
\def\el{{{\rm e}^--{\rm e}^+}}

\usepackage{graphicx}
\usepackage{mathrsfs}
\title[Dissipative advective accretion disc solutions]
{Dissipative advective accretion disc solutions with variable adiabatic index 
around black holes}
\author[Kumar \& Chattopadhyay]
{Rajiv Kumar$^{1}$, Indranil Chattopadhyay$^{1}$\thanks{E-mail: rajiv.k@aries.res.in (RK);
indra@aries.res.in (IC)}\\
$^{1}$Aryabhatta Research Institute of Observational Sciences (ARIES), Manora Peak, Nainital-263002, India\\
}
\begin{document}
\date{}
\maketitle
\label{firstpage}

\begin{abstract}
We investigated accretion on to black holes in presence of viscosity and cooling, by employing
an equation of state with variable adiabatic index and multi-species fluid.
We obtained the expression of generalized Bernoulli parameter which is a constant of motion
for an accretion flow in presence of viscosity and cooling. We obtained all possible
transonic solutions for a variety of boundary conditions, viscosity parameters and accretion rates.
We identified the solutions with their positions in the parameter space of generalized Bernoulli parameter
and the angular momentum on the horizon.
We showed that a shocked solution is more luminous than a shock-free one.
For particular energies and viscosity parameters,
we obtained accretion disc luminosities in the range of $10^{-4}-1.2$ times Eddington limit,
and the radiative efficiency seemed to increase with the mass accretion rate too.
We found steady state shock solutions even for high-viscosity parameters, high accretion rates, and
for wide range of composition of the flow, starting from purely electron-proton to lepton-dominated accretion flow.
However, similar to earlier studies of inviscid flow, accretion shock was not obtained for electron-positron
pair plasma.
\end{abstract}


\section {Introduction}
Observations of electromagnetic spectra and mass outflow in the form of jets
from microquasars and active galactic nuclei (AGN) are better explained as a consequence of
conversion of gravitational
energy released from matter falling into extreme compact objects like black holes.
AGN are supposed to harbour $10^{6-9}~M_{\odot}$ (where, $M_{\odot}$ is solar mass) black holes
at the centre, and microquasars harbour black holes of mass $\sim 10~M_{\odot}$ at the centre.
Microquasars moves from a `low hard
spectral state' (LHS) \ie when the accretion disc is radiatively in-efficient and the
electro-magnetic power maximizes in the higher energy power-law tail,
to the `high soft spectral state' (HSS) \ie when the disc is radiatively
efficient and the power maximizes in the lower, thermal part of the spectra
\citep{rm06}.
These two states are connected by a series of intermediate states.  
In fact, the time-scales in microquasars and AGN scale
with the mass of the central object \citep{mkkf06}, and shows that the basic physics close to the horizon
is similar. As a result, timing properties can be studied by observing the microquasars which are much faster varying
than the AGN. In particular, since the time-scales are large, AGN generally do not show state transition,
therefore, a closer look at the timing properties
of microquasars may
shed some light on when the transitions are expected for AGN. Microquasars also show quasi-periodic oscillation
(QPOs) in the hard power law photons \citep{rm06}, and in the out bursting sources, the QPO frequency grows
with the luminosity and as the spectral state moves from LHS to the intermediate states \citep{st09,ndmc12}. 
Moreover, black hole candidates (BHCs) are associated with bipolar jets \citep{b93,mr94}. Since
black holes have no atmosphere, therefore, jets have to originate from the accretion disc.
Observations also suggest that these jets originate within a few tens of Schwarzschild radii ($r_{\rm g}$)
from the horizon \citep{jbl99,detal12}.
Although from centres of galaxies jets have been observed either to exist 
or none at all, however, microquasars jets have been observed to evolve with the change in its spectral state
\citep{gfp03}. In other words, if we accept the conclusions of \citet{mkkf06}, then those centres of galaxies which
harbours central black holes but do not exhibit the presence of jet, should start generating jets as and when the
state of the accretion disc changes to jet generating mode, and vice versa.
Since accretion is the main driver that fuels AGN and microquasars,
so it is very important to understand the accretion physics in details.  

It is well known that matter enters the black hole with the speed of light or $c$ \citep{w72},
and at large distances away from the horizon
should have negligible infall velocity, therefore, matter accreting on to black holes are necessarily
transonic, or in other words, matter falling on to a black hole makes a subsonic to supersonic transition.
Moreover, the presence of marginally stable orbit ensures that matter entering a black hole must have sub-Keplerian
angular momentum.
These features are imposed by the inner boundary condition of the space-time geometry around the black hole. 
It implies that black hole accretion should have significant advection. 
The very first model for accretion on to black holes is the general relativistic version of 
Bondi flow \citep{b52,m72}, which satisfies all the above properties of black hole accretion.
But this model could not explain observed luminosities around BHCs, since Bondi flow is 
a radial flow whose infall time-scale is too small to produce the luminosities observed \citep{s73}.
To circumvent this problem, the Keplerian disc or Shakura-Sunyaev (SS) standard disc model was proposed by
\citet{ss73,nt73}.
In this model, the disc matter is rotation dominated with Keplerian angular momentum distribution and negligible
infall velocity. The matter is optically thick, geometrically thin and was  
successful in explaining thermal multicoloured blackbody part of the spectrum but could not produce non-thermal part 
of the spectrum. 
It was shown that a Comptonizing cloud of hot electrons or some kind of a `corona' is required to produce the
high-energy non-thermal photons \citep{st80}. The differences in these models arise due to the importance assigned to the
various terms
in equations of motion. For Bondi flow, no rotation was considered, and instead of energy equation polytropic
equation of state (EoS) was assumed. For SS disc the advection term and the pressure gradient term in the momentum balance
equation were ignored, and the heat generated by the viscous disc was assumed to be efficiently radiated away.
Although Bondi flow failed as a viable accretion model to 
explain the luminosities of BHCs, but the unique inner boundary condition of a space-time
around a black hole would ensure the nature of accretion solutions close to the horizon to be quasi-Bondi type
\ie  advection would be dominant close to the horizon. As a result, accretion models with significant advection
gained popularity. 
\citet{nkh97} developed a class of solutions called advection-dominated accretion flow (ADAF), characterized by
rotating flow with viscous dissipation, and becomes transonic close to the horizon. It was also shown that
such flow becomes self-similar at large distances away from the horizon. ADAF got unprecedented popularity
as the disc model to explain the observation. However, earlier
\citet{lt80} showed that rotating transonic
flow in the inviscid limit may harbour multiple sonic points. Such accreting
flows may undergo transient or steady shock transitions \citep{f87,c89,mlc94,msc96,mrc96}.
Presence of multiple sonic point and shocks has been shown to exist for dissipative accretion flows
as well \citep{c96,lmc98,cd07,d07,dc08,lcscbz08,lrc11}. Whether an accretion flow is smooth
or shocked depends on the outer boundary condition, and ADAF-type solution has been shown to be
a subset of general advective, viscous solutions \citep{lgy99,bdl08,kc13,kcm14}.
A shocked solution has a number of advantages. The extra thermal gradient force in the post-shock
flow can drive bipolar outflows \citep{mrc96,lmc98,dcnc01,cd07,kc13,kcm14}, and since shocks
occur typically at few tens of $r_{\rm g}$, a shocked disc naturally satisfies
the observational criteria that jets are launched closer to the horizon, and the entire disc do not
participate in the formation of jets. Moreover, \citet{kcm14} showed that jets from post-shock disc becomes
stronger as the spectral state moves from LHS to hard intermediate spectral states as is observed in microquasars.

Taking clue from the solutions of shocked disc, \citet{ct95} in a model solution
considered Keplerian flow along the equatorial plane and sub-Keplerian flow which flanks the Keplerian
disc from above and below. The sub-Keplerian flow may harbour shock, and the post-shock flow being hot and puffed up
will intercept a portion of the pre-shock disc photons and inverse-Comptonize them to produce the hard power-law tail,
producing the LHS.
Very weak shock or no shock means lack of hot electrons, and therefore cannot produce the hard power-law tail, \ie forms the
HSS. Surge in Keplerian accretion rate or the sub-Keplerian accretion rate can be either due to actual supply
at the outer boundary,
or by redistribution between Keplerian and sub-Keplerian matter
due to the change in microphysics which results in a change in viscosity parameter. Evidence of Keplerian and sub-Keplerian
matter has been confirmed observationally about a decade ago \citep{shms01,shs02,sds07}, and recently numerical
simulations have confirmed the two-component paradigm \citep{gc13}. Since post-shock disc is the seat of hard radiation,
so oscillating shocks \citep{msc96,lcscbz08,ny09,lrc11,dcnm14} will induce oscillations in hard radiations as well. Hence,
this could naturally explain the QPO. 
Therefore, entire advective regime which gives rise to smooth and shocked solutions can qualitatively explain three
broad observational features from BHCs like spectral states and its transition, jets, evolution of QPOs.

Most of the above work has been done with fixed adiabatic index ($\Gamma$) EoS. Since 
accretion flow at very large distance from the black hole is thermally
non-relativistic, \ie $\Theta(=kT/m_{{\rm e}^-}c^2)<1$, and close to it 
becomes relativistic or $\Theta\gg1$, therefore, $\Gamma$ should not be a fixed quantity in the flow but 
depend on temperature. In the definition of $\Theta$, $k$ is the Boltzmann constant, $T$ is the temperature
and $m_{{\rm e}^-}$ is electron rest mass. In fact, \citet{t48} showed that a fixed $\Gamma$ EoS should not be
used to obtain solutions of flow which covers non-relativistic to
relativistic temperature ranges. \citet{bm76} presented the first accretion solution using temperature-dependent
$\Gamma$ EoS around black holes. \citet{f87} showed the presence of accretion shock for a flow composed of electrons and
protons ($\ep$). \citet{c08} and \citet{cr09} showed that flow solutions around a black hole
depend on the composition of the flow, and that the steady state accretion solutions of electron-positron
pair plasma ($\el$) possesses flows with non-relativistic temperatures,
and as a consequence, do not form accretion shocks \citep{c08,cc11}.
Since the relativistic nature of EoS depends on the competition between rest energy and the thermal energy of the
flow, so it was found out that a flow becomes thermally the most relativistic if its  proton proportion
is $25-27\%$ of its electron number density \citep{cr09,cmggkr12}. And the maximum proportion
of bipolar outflow is generated when the composition parameter is in this range \citep{kscc13}.
Although accretion solutions with variable $\Gamma$ EoS around compact objects were obtained before, but as far as we are
aware, solutions of rotating, transonic flow with variable $\Gamma$ EoS which
are composed of baryons and leptons in presence of 
general heating and cooling have not been
attempted before. Since the thermal state of the flow and its advective properties ultimately determine
the radiative power output, the present attempt is important. Moreover, it would also be important to
determine the effect of accretion rate in shaping the nature of solutions and the luminosities generated
from such solutions. All these issues will be discussed in the subsequent sections.

In the next section, we introduce the governing equations and outline the assumptions, and, we describe the solution
procedure. In Section 3, we present the results, and finally in Section 4, we present the discussion and concluding remarks. 

\section {Assumptions and Equations of motion}
We consider stationary, viscous, rotating and axisymmetric accretion disc around a Schwarzschild black hole. 
For mathematical simplicity, space-time around the black hole is described by the Paczy\'nski-Wiita (PW)
pseudo-Newtonian potential \citep{pw80}. PW potential simplifies our calculations
while retaining all the essential qualitative features of strong gravity, and it is easier to incorporate
more complicated physics in the pseudo-Newtonian scheme. 

\subsection {Equations of motion}
The EoS of a multispecies flow is given by \citep{c08,cr09}
\be
{\bar e}=n_{{\rm e}^{-}}m_{{\rm e}^{-}}c^{2}f = \rho_{{\rm e}^{-}}c^2f = \frac{\rho c^2f}{K},\mbox{ where }
K ={[2 - \xi(1 - 1/{\eta})]}
\label{seed.eq}
\ee 
and,
$$
f = (2-\xi) \left[1 + \Theta \left(\frac {9\Theta + 3}{3\Theta + 2}\right)\right] + \xi \left[\frac{1}{\eta} + \Theta
\left(\frac {9\Theta + 3/\eta}{3\Theta + 2/\eta}\right)\right].
$$
Here $\xi=n_{{\rm p}^+}/n_{{\rm e}^-}$ is the ratio of proton to electron number density, and the ratio of electron
to proton mass is given by $\eta=m_{{\rm e}^-}/m_{{\rm p}^+}$. The approximate form of EoS of each species of the flow which follows a
relativistic
Maxwellian distribution was proposed by \citet{rcc06}. Following a different approach,
variable $\Gamma$ was considered for accretion
discs by assuming thermal equilibrium
between radiation and accreting flow \citep{md12}, but in our case no such assumption is required.
The charge neutrality is maintained by positron number density $n_{{\rm e}^+}$, so $n_{{\rm e}^-}
=n_{{\rm p}^+}+n_{{\rm e}^+}$.
The enthalpy is given by 
\be
h = \frac{(\bar e + p)}{\rho}  =  \frac{f}{K} + \frac{2\Theta}{K}.
\label{enth.eq}
\ee
where, $p=2n_{{\rm e}^-}kT$ is the isotropic gas pressure, $T$ being the local temperature, and $\Theta=kT/(m_{{\rm e}^-}c^2)$
is the ratio of thermal energy and the rest energy of the electron.
The expression of polytropic index and adiabatic index is given by
\be
N=\frac{1}{2}\frac{{\rm d}f}{{\rm d}\Theta}; ~~ \Gamma=1+\frac{1}{N}
\label{poly.eq}
\ee

We now employ the units of length, time and
velocity as $r_{\rm g}=2GM/c^2$, $r_{\rm g}/c=2GM/c^3$ and $c$, respectively,
where $M$ is the mass
of the black hole is the unit of mass, and $G$ is the gravitational constant.
In the rest of the paper, all the variables and equations are expressed in the above-mentioned unit system.
We now present steady state equations of motion in the advective domain and in presence of viscosity
and cooling processes. 
The radial component of momentum balance equation is given by
\be
\vel \frac{d\vel}{dx} + \frac{1}{\rho} \frac{dp}{dx} - {\frac {\lambda^2}{x^3}}   + \frac{1}{2(x-1)^2} = 0.
\label{rme.eq}
\ee
Here $\lambda$, $\vel$, and $x$ are the specific angular momentum, the radial bulk velocity, and the radial coordinate
in the units described above. 
The azimuthal component of momentum balance equation or, angular momentum distribution equation is given by,
\be
\frac{d\lambda}{dx}+\frac{1}{\Sigma \vel x}\frac{d(x^2 W_{x \phi})}{dx}=0,
\label{angm.eq}
\ee
The viscous stress is given by
$W_{x\phi}=\varpi x \frac{d\Omega}{dx}$, and the dynamical viscosity parameter
$\varpi=\Sigma \nu=\rho H (\alpha a^2)/(\Gamma \Omega_{\rm K})$, where,
$\nu$, $\alpha$ and $\Omega_{\rm K}$ are the kinematic viscosity parameter, Shakura-Sunyaev viscosity
parameter and the Keplerian angular velocity, respectively.
The mass conservation equation is given by,
\be
\dot{M}=2 \pi \Sigma \vel x,
\label{mf.eq}
\ee
where, $\Sigma=2\rho H$ is vertically integrated density of the flow. 
The disc matter is in hydrostatic equilibrium in the vertical direction, the half height of the disc is 
given by
\be
H=2\sqrt{\frac{\Theta x}{K}}(x-1).
\label{hh.eq}
\ee
The entropy generation equation or the first law of thermodynamics is given by
\be
\Sigma \vel \left[\frac{de}{dx} - \frac{p} {\rho^{2}} \frac{d\rho}{dx} 
+\frac{Q^+}{\vel}-\frac{Q^-}{\vel}\right] = 0,
\label{cee.eq}
\ee 
where ${e}={\bar e}/\rho=f/K$ is specific energy density of the flow. Here, $Q^{\pm}$ are the viscous heating
and the total cooling, respectively.
Here $Q^+=\frac{W_{x\phi}^2}{\eta \Sigma}$ and $Q^-=\Lambda^-/\Sigma$.
The cooling term is given by $Q^-=\chi {\mathscr F}(x_{\rm s})(Q_{\rm S}^-+Q_{\rm B}^-)$,
where $\chi$ is the cooling parameter, such that the cooling will be turned off by putting $\chi=0$, or
will be turned on if $\chi=1$. Here, ${\mathscr F}(x_{\rm s})$ is the Comptonization parameter fitting function,
which is calculated once the accretion shock is obtained and has been presented recently by \citet{kcm14}.
We assume ${\mathscr F}(x_{\rm s})$ to be generic.
The form of this analytical function is given by
\be
{\mathscr F}(x_{\rm s}) = −0.659234 + 0.127851x_{\rm s}-0.00043x_{\rm s}^2-1.13\times10^{-6} x_{\rm s}^3,
\label{compto.eq}
\ee
where, $x_{\rm s}$ is the location
of shock, and ${\mathscr F}(x_{\rm s})\sim 1$ when stable shock solution is not found.
The cooling term also contains synchrotron emissivity or $\Lambda_{\rm S}^-$ \citep{st83}, and bremsstrahlung 
emissivity or $\Lambda_{\rm B}^-$ \citep{rl79,s82} and are defined as
\be 
Q_{\rm S}^-=\Lambda_{\rm S}^-/\Sigma=\frac{S_0 \Theta_{\rm e}^{3}}{\vel\sqrt{\Theta x^3}(x-1)} ~~
\mbox{and} ~~
Q_{\rm B}^-=\Lambda_{\rm B}^-/\Sigma=\frac{B_0\sqrt{\Theta_{\rm e}}}{\vel\sqrt{\Theta x^3}(x-1)},
\label{sync.eq}
\ee 
where,
$$
S_0=\frac{16\times1.44\times10^{17}}{3}\frac{e^4\beta\dot{m}}{m_{\rm e}^3c^3K^{3/2}}\frac{1}{GM_{\odot}}
$$
and
$$
B_0=\frac{1.44\times10^{17}K_{\rm ep}\xi(2-\xi)\dot{m}}{16\pi m_{\rm e}^2K^{3/2}}\frac{1}{GM_{\odot}c^2}
$$
where $K_{\rm ep}=32 m_{\rm e}c^3r_{\rm e}^2
\alpha_{\rm f} \sqrt{(2/\pi)}/3=1.2135\times10^{-22}$, $\dot{m}={\dot M}/{\dot M}_{\rm Edd}$
is the accretion rate in units of Eddington rate,
the fine structure constant is given by
$\alpha_{\rm f}=1/137.036$ and classical electron radius $r_{\rm e}=2.81794\times10^{-13}$cm.
Here we have considered ${\dot M}_{\rm Edd}=1.44\times10^{17} M/M_{\odot}$.
The magnetic field is stochastic and is assumed to be in total or partial equipartition
with the gas pressure. The ratio between magnetic and gas pressure is $\beta=B^2/(8\pi p)$,
such that $0\leq \beta \leq 1$. 

Integrating equation (\ref{rme.eq}) with the help of eqs. (\ref{angm.eq} --- \ref{cee.eq}), we get
\be
\varepsilon=\frac{\vel^2}{2}+ h -\frac{\lambda^2}{2x^2}+\frac{\lambda \lambda_0}{x^2}-\zeta-\frac{1}{2(x-1)},
\label{ge.eq}
\ee
where $\zeta={\int{\frac{\Lambda^-}{\Sigma \vel}}}dx=\chi \int {\mathscr F}(x_{\rm s})
\left(({\Lambda_{\rm S}^-})/({\Sigma \vel})+
({\Lambda_{\rm B}^-})/({\Sigma \vel})\right)dx=\zeta_{\rm S}+\zeta_{\rm B}$. 
Here $\varepsilon$ is a constant of motion in presence of viscosity and cooling, and we call it the generalized
Bernoulli constant, and is also the specific energy of the flow.
If we ignore cooling processes in the equation (\ref{cee.eq}) then integral
form of equations (\ref{rme.eq}), (\ref{angm.eq}) --- (\ref{cee.eq}) becomes
\be
E=\frac{\vel^2}{2}+ h -\frac{\lambda^2}{2x^2}+\frac{\lambda \lambda_0}{x^2}-\frac{1}{2(x-1)}.
\label{spe.eq}
\ee
This is known as grand specific energy \citep{gl04,kc13} and is a constant of motion in presence of
viscosity but not in presence of cooling. And for non-dissipative flow
${\varepsilon}\rightarrow {\cal E}=0.5\vel^2+h+\lambda^2/(2x^2)-0.5/(x-1)$ which is the canonical definition
of Bernoulli parameter. 

Integrating equation (\ref{cee.eq}) by putting $Q^+=Q^-=0$, 
we get 
\be
\rho={\cal{K}} {\rm exp}(k_3) \Theta^{3/2}(3\Theta+2)^{k_1}(3\Theta+2/\eta)^{k_2},
\label{aes.eq}
\ee
where, $k_1=3(2-\xi)/4, k_2=3\xi/4$, and $k_3=(f-K)/(2\Theta)$.
This is the adiabatic EoS for multispecies fluids \citep{kscc13} and ${\cal K}$ is the constant of entropy.
Using equations (\ref{mf.eq}) and (\ref{aes.eq}), we can define entropy-accretion rate ($\md$) as
\be
\md = \frac{\dot{M}}{4\pi \cal{K}}= \vel H x {\rm exp}(k_3) \Theta^{3/2}(3\Theta+2)^{k_1}(3\Theta+2/\eta)^{k_2},
\label{enar.eq}
\ee
where $\md$ is also constant for inviscid multispecies relativistic flows.

Integrating equation (\ref{angm.eq}) with the help of equation (\ref{mf.eq}) and using expression of 
$W_{x\phi}$, we get
\be
\frac{d\Omega}{dx}=-\frac{\Gamma\vel\Omega_{\rm K}(\lambda-\lambda_0)}{\alpha a^2 x^2},
\label{dom.eq}
\ee
where, $\lambda_0$ is the specific angular momentum at the horizon obtained by considering vanishing torque
at the event horizon \citep{w72} and $\Omega_{\rm K}^2(x)=1/(2x(x-1)^2)$.
Since
$\lambda=x^2\Omega$, derivative of $\lambda$ with respective to $x$ is written as
\be
\frac{d\lambda}{dx}=2x\Omega+x^2\frac{d\Omega}{dx}.
\label{dlmd.eq}
\ee

Using equations (\ref{mf.eq}) and (\ref{hh.eq}) in equation (\ref{cee.eq}), we get
\be
\frac{d\Theta}{dx} = -\frac{2\Theta}{2N+1} \left[\frac{1}{\vel} \frac{d\vel}{dx} + 
\frac{5x-3}{2x(x-1)}\right]-\frac{\nu x^2 K}{(2N+1)\vel}\left(\frac{d\Omega}{dx}\right)^2+
\frac{K{\mathscr F}(x_{\rm s})}{(2N+1)\vel}[Q_{\rm S}^-+Q_{\rm B}^-].
\label{dth.eq}
\ee
Using equations (\ref{enth.eq}) and (\ref{dth.eq}) in equation (\ref{rme.eq}), we get  
\be
\frac{d\vel}{dx} = \frac {a^2 [{\frac{2N}{2N+1}} {\frac{5x-3}{2x(x-1)}}] +
\frac{\nu x^2}{\vel(2N+1)}\left(\frac{d\Omega}{dx} \right)^2-\frac{{\mathscr F}(x_{\rm s})}{\vel(2N+1)}[Q_{\rm S}^-
+Q_{\rm B}^-] +
{\frac {\lambda^2}{x^3}} - {\frac {1}{2(x-1)^2}}} {\vel - {\frac{a^2} { \vel}} [{\frac {2N} 
{2N+1}]}},  
\label{dv.eq}
\ee
where, $a$ is the adiabatic sound speed and is defined as
$a^{2} = 2 \Theta \Gamma/K$. To find a complete set of solution we have to integrate eqs. (\ref{dlmd.eq}---\ref{dv.eq}),
with the help of equation (\ref{dom.eq}). These equations are solved by specifying the flow parameters, namely, $\varepsilon$,
$\lambda_0$, $\alpha$ and ${\dot m}$. In addition, the lack of knowledge of the exact nature of black hole magnetosphere,
influences us to assume the value of $\beta$ to evaluate $Q^-_{\rm S}$.
In the case of black hole accretion, the presence of horizon imposes
at least one sonic or critical point ($x_{\rm c}$). Although, eqs. (\ref{dlmd.eq})---(\ref{dv.eq}) can be solved
once the flow parameters are supplied, but for dissipative flow the location of $x_{\rm c}$ or the number
of $x_{\rm c}$s, is not known a priori. 

\subsubsection{Critical point conditions}
Since accretion on to a black hole is transonic, therefore, at some point the denominator
${\cal D}$ of equation (\ref{dv.eq}) becomes zero, and to maintain the well behaved nature of d$\vel/$d$x$
the numerator ${\cal N}$ also goes to zero. This point where d$\vel/$d$x={\cal{N}/\cal{D}}\longrightarrow0/0$
is called the critical point or the sonic point ($x_{\rm c}$) of the flow. This also gives us the critical point condition,
\be
M_{\rm c}^2=\frac{\vel_{\rm c}^2}{a_{\rm c}^2}=\frac{2}{\Gamma_{\rm c}+1}
\label{dc.eq}
\ee
and
\be
 \left[\frac{(5x_{\rm c}-3)M_{\rm c}^2}{2x_{\rm c}(x_{\rm c}-1)}\right]a_{\rm c}^2 + 
\frac{\Gamma_{\rm c} M_{\rm c} \Omega_{\rm K}(\lambda_c-\lambda_0)^2}{\alpha a_{\rm c} x_{\rm c}^2(2N_{\rm c}+1)} -
\frac{{\mathscr F}(x_{\rm s})[Q_{{\rm S_c}}^-+Q_{{\rm B_c}}^-]}{\vel_{\rm c}(2N_{\rm c}+1)} +
{\frac {\lambda_{\rm c}^2}{x_{\rm c}^3}} - {\frac {1}{2(x_{\rm c}-1)^2}}=0,
\label{nc.eq}
\ee
where $M_{\rm c}, \vel_{\rm c}, a_{\rm c}, \Gamma_{\rm c}, N_{\rm c},~\mbox{and } \lambda_{\rm c}$
are the Mach number, the bulk velocity, 
the sound speed, the adiabatic index, the polytropic index and the specific angular momentum
at the critical point $x_{\rm c}$, respectively. The bulk velocity gradient at the critical
point is calculated by l$^{\prime}$Hospital rule and is given by
\be
\left(\frac{d\vel}{dx}\right)_{\rm c}=\left(\frac{d{\cal{N}}/dx}{d{\cal{D}}/dx} \right)_{x=x_{\rm c}}.
\label{dvc.eq}
\ee
Equations (\ref{dc.eq}---\ref{dvc.eq}) give the analytical critical point conditions,
which are used to obtain the critical point of the flow.

\subsubsection{Shock conditions}
In the domain of multiple sonic points, supersonic matter through outer sonic point, may be slowed down due to the centrifugal
barrier at $x\lsim$few$\times10~r_{\rm g}$. This may act as a barrier to the supersonic matter following the
slowed down matter. If the barrier is strong enough it may produce a centrifugal barrier mediated shock
transition.
The shock conditions are obtained from conservation of mass, momentum and energy fluxes across the discontinuity
\citep{ll59}. The general, compact, and conserved form of the fluid equations are, 
$$
\partial_t(q)+\partial_x(F_q)=0,
$$
where, $q$s are the conserved quantities and $F_q$ are corresponding fluxes. We now impose the conditions that $v_z=0$ and $\partial/\partial \phi
=\partial/\partial z=0$, and only the $x-\phi$ component of the viscous stress is the most dominant.
Assuming hydrostatic balance in the vertical direction we obtain the integrated form of the mass flux
($F_{\rm mass}$), the
radial momentum flux ($F_{x-{\rm mom}}$), the azimuthal momentum flux ($F_{\phi-{\rm mom}}$),
and the energy flux ($F_{\rm energy}$) in the
radial direction, and are given by, 
\be
F_{\rm mass}={\dot M}, ~ F_{x-{\rm mom}}=(W+\Sigma \vel^2),~ F_{\phi-{\rm mom}}={\dot J}=
\dot{M}\lambda+x^2W_{x\phi}~{\rm and}~F_{\rm energy}={\dot M}(\varepsilon-\Phi),
\label{fluxs.eq}
\ee
where, $\Phi$ is the gravitational potential. It is quite interesting to see that, the 
mass flux along the radial direction is ${\dot M}$, the momentum flux in the radial direction
is the sum of the thermal pressure and the ram pressure, in the azimuthal direction the momentum flux
is ${\dot J}$ or the angular momentum flux, and the
energy flux is related to the generalized Bernoulli parameter $\varepsilon$.
After some straight forward algebra, we find
the generalized version of the non-dissipative
shock condition in presence of viscosity and cooling, and is given by
\be 
\dot{M}_+= \dot{M}_- ,
\label{mfs.eq}
\ee

\be 
W_+ +\Sigma_+\vel_+^2=W_- +\Sigma_-\vel_-^2 ,
\label{momf.eq}
\ee
\be
\dot{J}_+=\dot{J}_-,
\label{angf.eq}
\ee
and
\be 
\varepsilon_+= \varepsilon_-,
\label{enf.eq}
\ee
where subscripts minus(-) and plus(+) denote the quantities of supersonic and subsonic branches across
the shock in a black hole accretion flow, respectively.
In the inviscid and adiabatic limit if we put $\alpha=Q^-=0$, we retrieve the original Rankine Hugoniot
shock conditions, \ie equation (\ref{angf.eq}) is redundant and equation (\ref{enf.eq}) reduces to the conservation
of the canonical Bernoulli parameter across the shock \citep{ll59,c89}. Various authors have used various forms
of shock conditions for viscous flow, in the literature. The shocked disc solution with viscosity was obtained by
\citet{c96}
and later by \citep{gl04,cd07,d07,dc08},
by choosing the viscous stress to be proportional to the total pressure, which made equation (\ref{angf.eq})
redundant. \citet{bdl08} on the other hand, used isothermal condition instead of equation (\ref{enf.eq}).
In \citet{kc13} and \citet{kcm14}, cooling was ignored, so equation (\ref{enf.eq}) was replaced by the conservation of $E$
across the shock front.
Since, equations (\ref{mfs.eq}-\ref{enf.eq}) do not explicitly depend on the form
of $h$, and viscosity and cooling processes have been considered,
therefore this form of shock condition is the most general form of Rankine-Hugoniot type
shock conditions obtained by strict conservation of fluxes of the equations of motion
by following the prescription laid down by \citet{ll59}. 

In \citet{kc13}, we have discussed various types of dissipative shocks, in which the fraction of
thermal energy dissipated was supplied as a parameter.
In this paper, we would discuss a special form of dissipative shock by replacing equation (\ref{enf.eq})
with
\be
E_+=E_-.
\label{isoth.eq}
\ee

Using equations (\ref{mfs.eq}-\ref{enf.eq}), the supersonic branch radial
velocity, temperature, 
and the angular momentum can be obtained from the post-shock quantities and vice versa,
\be
\vel_-^2-2\left(c_1-h_-+\frac{\lambda_-^2}{2x_{\rm s}^2}-\frac{\lambda_-\lambda_0}{x_{\rm s}^2}+\zeta_-\right)=0,
\lambda_-=\lambda_0+\frac{c_2 a_-^2}{\Gamma_-\vel_-}
\\~~\mbox{and}\\~~ \Theta_-=\frac{K}{2}\left(c_0 \vel_--\vel_-^2\right),
\label{suq.eq}
\ee
where $c_0=[2\Theta_+/K+\vel_+^2]/\vel_+$, $c_1=\vel_+^2/2+h_+-\lambda_+^2/(2x_{\rm s}^2)+
\lambda_+\lambda_0/x_{\rm s}^2-\zeta_+$,~ $\zeta_+ = {\zeta_{\rm S}}_+ + {\zeta_{\rm B}}_+$,
$c_2=\Gamma_+\vel_+(\lambda_+-\lambda_0)/(a^2_+)$
and 
$\zeta_-=(f_{\Theta_{\rm e}}^3 {\zeta_S}_++f_{\Theta_{\rm e}}^{1/2}{\zeta_{\rm B}}_+)/(f_{\vel}^2 f_{\Theta}^{1/2})$.
Moreover, $f_{\Theta_{\rm e}}={\Theta_{\rm e}}_-/{\Theta_{\rm e}}_+$,
$f_{\vel}=\vel_-/\vel_+$ and $f_{\Theta}=\Theta_-/\Theta_+$.
All three quantities ($\vel_-, \Theta_-$ and $\lambda_-$) in equation \ref{suq.eq} are obtained simultaneously 
in terms of post-shock quantities which eventually gives us the shock location $x_{\rm s}$.

For dissipative shocks, we use equations (\ref{mfs.eq})-(\ref{angf.eq}) and (\ref{isoth.eq}) to relate the post-shock and
the pre-shock
quantities, and they are
\be
\vel_-^2-2\left(c_3-h_-+\frac{\lambda_-^2}{2x_{\rm s}^2}-\frac{\lambda_-\lambda_0}{x_{\rm s}^2}\right)=0,
\\~~\lambda_-=\lambda_0+\frac{c_2 a_-^2}{\Gamma_-\vel_-}
\\~~\mbox{and}\\~~ \Theta_-=\frac{K}{2}(c_0 \vel_--\vel_-^2),
\label{iso.eq}
\ee
where, $c_3=\vel_+^2/2+h_+-\lambda_+^2/(2x_{\rm s}^2)+\lambda_+\lambda_0/x_{\rm s}^2$.

\subsection{Solution Procedure}
As has been discussed in Section 2.1, the equations of motions, \ie equations (\ref{rme.eq})-(\ref{mf.eq})\
and (\ref{cee.eq}),
can be simplified as
gradients of $\vel$ (equation \ref{dv.eq}), $\lambda$ (equation \ref{dlmd.eq}) and $\Theta$ (equation
\ref{dth.eq}).
Since black hole accretion is necessarily transonic, and the sonic point for dissipative flow is not
known a priori, so as the first step to find
complete solutions, one need to find a method to compute the location of the sonic point.
Moreover, black hole accretion may harbour at least one sonic point, but depending upon
boundary conditions the accretion solution may possess two physical sonic points, and one unphysical one.
One of the physical sonic point is closer to the horizon and is called inner sonic point ($x_{\rm ci}$),
the one further out is called outer sonic point ($x_{\rm co}$) and the unphysical sonic point is situated
in between $x_{\rm ci}$ and $x_{\rm co}$ and is called the middle sonic point $x_{\rm cm}$. The middle sonic point
is O type for inviscid, adiabatic flow but is spiral type for dissipative flow. In the domain of a single sonic
point, depending on boundary conditions, the single physical sonic point may be located closer to the black
hole or may be located far away from the black hole.

\subsubsection{Method to find the sonic point}
In principle, once the flow parameters like $\varepsilon$, $\lambda_0$, $\alpha$, ${\dot m}$ and $\beta$
are supplied one should be able to obtain the accretion solution by integrating equations (\ref{dlmd.eq})-(\ref{dv.eq}).
Since, at least one sonic point exists for black hole accretion, and we do not know the location of
the sonic point or the value of $\lambda_{\rm c}$
before hand, we have to find sonic point and $\lambda_{\rm c}$  by iteration method.
Moreover, even though all the flow parameters on the horizon
are known, but because of the coordinate singularity on the horizon, we cannot start the integration from the
horizon itself. Therefore, we consider the following steps to obtain the sonic points.

{\it Step 1}. We estimate the asymptotic behaviour of the flow variables at $x_{\rm in}$ a radial coordinate
very close to the horizon, and use it as the starting point of our integration.
This method was successfully employed by \citet{bl03} for fixed $\Gamma$ EoS, here we implement the same method
for an EoS described by equation (\ref{seed.eq}). In addition we do not need to make an explicit assumption of
free fall close to the horizon.
Equation (\ref{dlmd.eq}) is a first-order differential equation, so we
can expand it by Frobenius expansion for $\lambda(x)$ about $r_{\rm g}$ \citep{bl03}.
\be
\lambda(x)=\lambda_0+{\cal{B}}(x-r_{\rm g})^{\delta}, ~~~~~~x \longrightarrow r_{\rm g},
\label{asyl.eq}
\ee
where constants ${\cal B}$, $\delta$ are to be determined. 
Combining equations (\ref{dom.eq} and \ref{dlmd.eq}), we obtain
\be
\frac{d\lambda}{dx}=2x\Omega -\frac{\Gamma\vel\Omega_{\rm K}(\lambda-\lambda_0)}{\alpha a^2}
\label{dlmd2.eq}
\ee
Using equation (\ref{asyl.eq}) in equation (\ref{dlmd2.eq}), in the limit $x\rightarrow r_{\rm g}$ we obtain,
\be
\lim_{x \to r_{\rm g}}\frac{\Gamma \vel \Omega_{\rm K}}{\alpha a^2}{\cal B}(x-r_{\rm g})^{\delta}
=\frac{2\lambda_0}{r_{\rm g}}-\lambda^{\prime}_0
\label{dlmd3.eq}
\ee
We replace $\vel$ between equations (\ref{dlmd3.eq}) and (\ref{enar.eq}), use the expression of $\Omega_{\rm K}$, and
the definition of $H$ (equation \ref{hh.eq}), we obtain,
\be
\lim_{x \to r_{\rm g}}\frac{\Gamma \md K^{1/2}}{\alpha a^2 2{\sqrt{2\Theta}}x^2(x-r_{\rm g})^2(\rho/{\cal K})}
{\cal B}(x-r_{\rm g})^{\delta}=\frac{2\lambda_0}{r_{\rm g}}-\lambda^{\prime}_0,
\label{dlmd4.eq}
\ee
where $\rho/{\cal K}$ is given by equation (\ref{aes.eq}).
A constant value warrants that net exponent of $(x-r_{\rm g})$ goes to zero,
\ie $\delta=2$. Since by equation (\ref{asyl.eq}), $\lambda\rightarrow \lambda_0$ as $x\rightarrow r_{\rm g}$,
therefore we have
\be
{\cal B}=4\alpha\lambda r_{\rm g}^2\frac{\sqrt{2\Theta}}{\Gamma\sqrt{K}\md}
{\rm exp}(k_3) \Theta^{3/2}(3\Theta+2)^{k_1}(3\Theta+2/\eta)^{k_2}.
\label{bexp.eq}
\ee
We now combine equations (\ref{asyl.eq}) and (\ref{bexp.eq}), and plug them in equation (\ref{ge.eq}).
In addition, $\vel$ 
in equation (\ref{ge.eq}) is 
expressed in terms of $\md$, hence we obtain a polynomial in $\Theta$.
Now, providing the parameters $\varepsilon, \lambda_0, \alpha, \xi,~\md~\mbox{and }{\cal F}(x_{\rm s})=1$, we
solve for $\Theta$ at $x_{\rm in}$. Once $\Theta_{\rm in}$ is obtained, $a_{\rm in}$, $\vel_{\rm in}$ and
$\lambda_{\rm in}$ are easily obtained.

{\it Step 2}. We now integrate equations (\ref{dlmd.eq}) --- (\ref{dv.eq}) outward from $x_{\rm in}$, with the asymptotic flow variables,
and simultaneously check the critical point conditions (equations \ref{dc.eq}-
\ref{dvc.eq}).

{\it Step 3}. Initial solution with an initial guess value of $\md$ will, in all probability, not be a transonic solution.
We change the value of $\md$ keeping other flow parameters same, and then again recalculate the asymptotic
flow variables at $x_{\rm in}$. With the new set of $\vel_{\rm in},~ \Theta_{\rm in},~ \lambda_{\rm in}$
we now follow Step 2, until the critical point conditions 
are satisfied, and obtain the sonic point or critical point ($x_{\rm c}$) of the flow.

{\it Step 4}. Once the we get value of $x_{\rm c}$ and the corresponding $\lambda_{\rm c}$, $\vel_{\rm c}$, $\Theta_{\rm c}$,
we compute $(d\vel/dx)_{x_{\rm c}}$ and $(d\Theta/dx)_{x_{\rm c}}$.

{\it Step 5}. Now the integration is continued outwards from $x_{\rm c} \rightarrow \infty$ to obtain the global solution.

The global solution might be smooth passing through one $x_{\rm c}$, or may be discontinuous harbouring shocks
and therefore passing through both the outer ($x_{\rm co}$) and the inner sonic ($x_{\rm ci}$) points.
One set of solution is also possible when the parameter space posses multiple sonic points,
but steady state shock conditions are not satisfied.

\subsubsection{To find the shock:}
As the equations of motion are integrated from $x_{\rm c}$ outward, we check either for the non-dissipative shock condition
(equations \ref{mfs.eq}-\ref{enf.eq}) or the dissipative shock conditions (equations \ref{mfs.eq}-\ref{angf.eq},
\ref{isoth.eq})
we calculate the supersonic branch quantities $\vel_-,~\Theta_-,~\mbox{and}~\lambda_-$ at the tentative jump radius
${\bar x}_{\rm s}$. Using these variables and ${\bar x}_{\rm s}$ as the starting point, we solve the equations of motion
to find out the outer sonic point of $x_{\rm co}$ by checking the sonic point conditions
(\ref{dc.eq}-\ref{dvc.eq}) iteratively. Once $x_{\rm co}$ is determined, then the corresponding ${\bar x}_{\rm s}$ 
is the tentative shock location. Now supplying ${\bar x}_s$ in equation (\ref{compto.eq}), we
find the Comptonization factor ${\mathscr F}({\bar x}_{\rm s})$ and update $Q^-$.
With this new cooling, we recalculate the shock location
again by retracing the steps suggested in Sections 2.2.1 and 2.2.2.
Once the shock location converges to a value $x_{\rm s}$, we have a self consistent shocked accretion solution.

 \begin{figure}
 \begin{center}
  \includegraphics[width=12.cm]{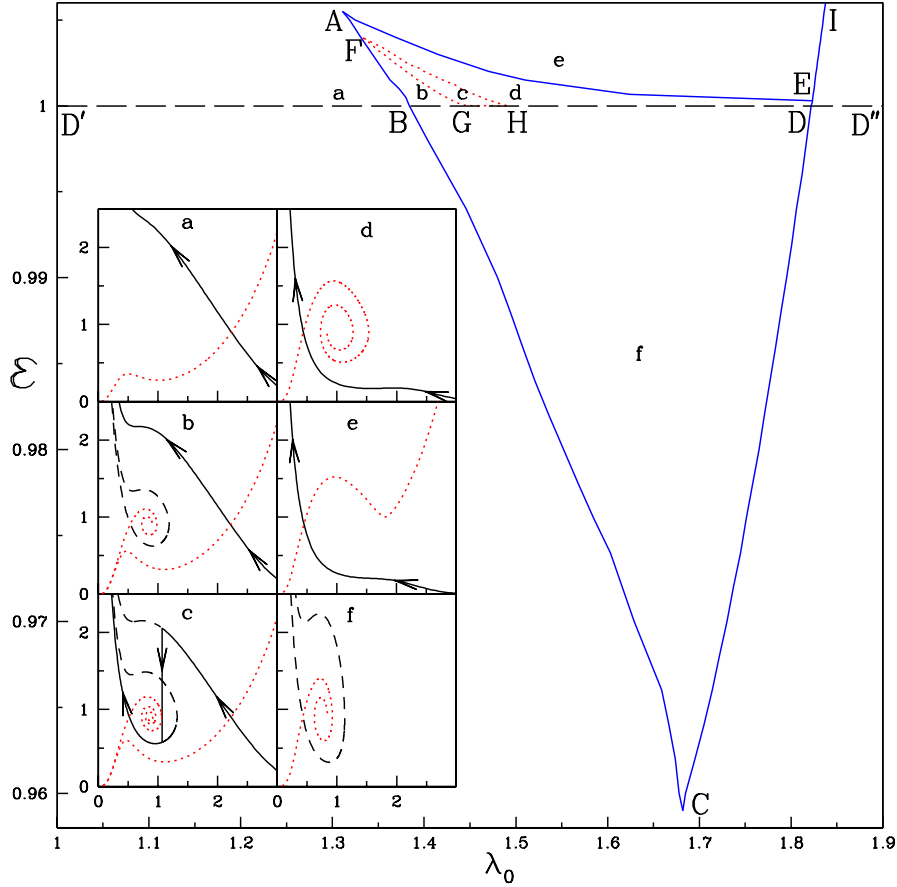}
\caption{Division of the parameter space in $\varepsilon - \lambda_0$ according to number of critical points 
and representative accretion solutions. Area ABDEA and BCDB has three and two critical points, respectively.
Area outside the bounded region D$^{\prime}$BAEI and $\varepsilon>1$ has single critical point, but for 
regions outside D$^{\prime}$BCDEI and $\varepsilon<1$  
no critical points exit.
Inset panels labelled as a, b, c, d, e, and f present Mach number $M=\vel/a$ versus log$(x)$, corresponding to
the $\varepsilon,~\lambda_0$ values at the locations marked in the parameter space.
The dotted region FGHF is the shock parameter space, and in panel (c) the vertical jump shows the position of
the shock. Accretion solutions are represented by solid (online black) curve. This parameter
space and the associated solutions are for $\alpha=0.05,~ \beta=0.01~\mbox{and}~\dot{m}=0.1$.}
\end{center}
\label{lab:fig1}
\end{figure}

\section{Results}

The accretion solutions are characterized by the following flow parameters: the generalized Bernoulli parameter
$\varepsilon$, and ${\dot m}$ which are constants of motion. Furthermore, $\lambda_0$ or the angular momentum
on the horizon is a constant of integration, and viscosity parameter $\alpha$ are the two more parameters.
On the top of that, $\xi$ the composition fraction determines the flow composition and therefore the EoS,
and $\beta$ controls the synchrotron emission, by estimating the magnetic energy. It is to be understood that
the equations of motion (equations \ref{mfs.eq}-\ref{enf.eq}) are not over determined, because ${\dot m}$ and $\beta$
together controls the cooling processes.
It is to be remembered that
supplying the inner boundary condition ($\varepsilon,~\lambda_0,~
{\dot m}$) to determine the sonic points in presence of $\alpha$, $\xi$ and $\beta$
is equivalent to, supplying the outer boundary condition ($\varepsilon,~\lambda_{\rm inj},~{\dot m}$),
where $\lambda_{\rm inj}$ is the specific angular momentum at the outer boundary. The outer boundary of the disc
is symbolized by $x_{\rm inj}$.
Following eqs. (\ref{sync.eq}), the total surface luminosity of the disc is given by
\be
\mathscr{L}_{\rm t}=4\pi\int {\mathscr F}(x_{\rm s})(Q_{\rm S}^-+Q_{\rm B}^-)\Sigma x{\rm d}x, ~~\mbox{and } \ell=\mathscr{L}_{\rm t}/
L_{\rm Edd},
\ee
where $\mathscr{L}_t$ is the total luminosity and $\ell$ is the dimensionless luminosity in units 
of $L_{\rm Edd}\approx1.3\times10^{38}(M/M_{\odot})$. The relation between Eddington accretion rate
and Eddington limit is ${\dot M}_{\rm Edd}=L_{\rm Edd}/c^2$.

\subsection{All possible transonic accretion solutions for $\ep$ flow}

\begin{figure}
 \begin{center}
  \includegraphics[width=12.cm]{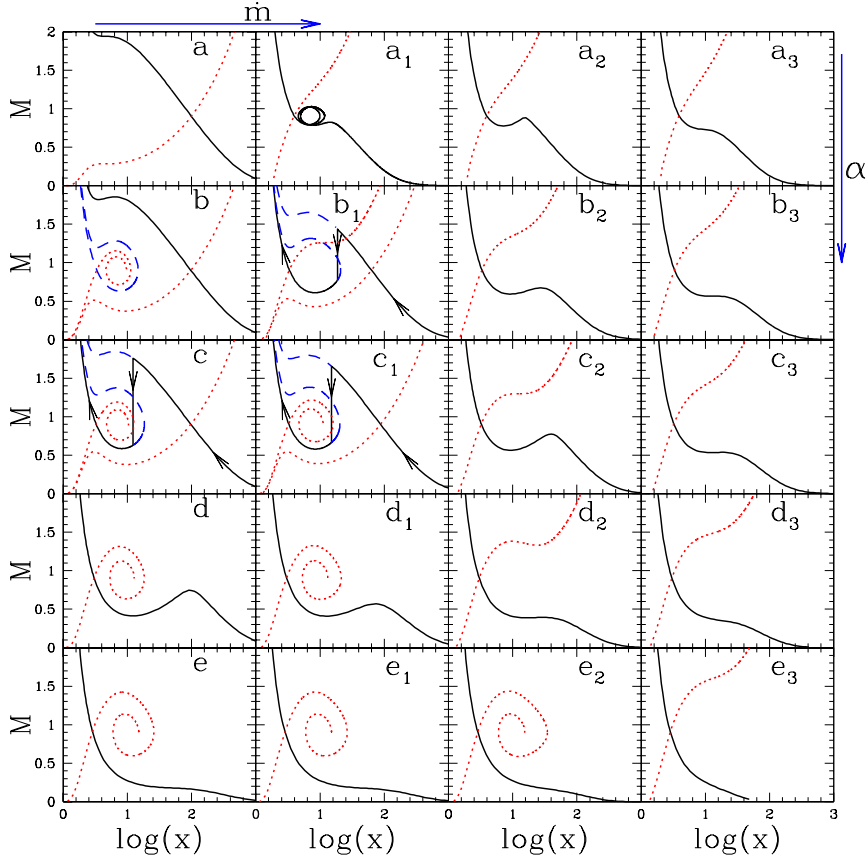}
  \caption{Effect of accretion rate and the viscosity parameter on shock-free inviscid flow.
The Mach number $M$ are plotted with log$(x)$ in all the panels,
for parameters $\varepsilon=1.001,~\lambda_0=1.46$, $\beta=0.1$ and $\xi=1.0$.
(a) Shock free solution with only $x_{\rm co}$ in the inviscid limit. Vertically down $\downarrow$
viscosity increases
\eg (a - e), 
$\alpha=0.0,~ 0.025,~ 0.03,~ 0.048$ and $0.065$, respectively, and $\chi=0.0$. Left to right,
${\dot m}$ increases. For panels $a_1 - a_3$, we have cooling \ie
$\chi=1.0$ and the accretion rates are ${\dot m}=3.5, 4.0$ and $5.0$.
For panels $b_1 - b_3$, $\chi=1.0$ and ${\dot m}=0.7,~1.4$ and $2.1$.
For panels $c_1 - c_3$, $\chi=1.0$ and ${\dot m}=0.2,~0.8$ and $1.6$.
For panels $d_1- d_3$, $\chi=1.0$ and ${\dot m}=0.1,~0.5$ and $1.0$.
For panels $e_1 - e_3$, $\chi=1.0$ and ${\dot m}= 0.01, 0.1$ and $1.0.$ Wind
type solutions are plotted as dotted curve (red online) and dashed (blue) part of the transonic solution
which
is not followed by the flow.}
 \end{center}
\label{lab:fig2}
\end{figure}

In Fig. 1, we present the full $\varepsilon-\lambda_0$ parameter space, and all possible
accretion solutions, corresponding to $\xi=1.0$ or $\ep$ flow,
$\alpha=0.05,~ \beta=0.01~\mbox{and}~\dot{m}=0.1$.
The solutions or the Mach number distributions $M~(=\vel/a)$ for various parameters
are plotted in the inset. In Fig. 1, the inset panel labelled `a' presents
$M$ versus log$(x)$, corresponding to location `a'
in the energy-angular momentum parameter space for coordinates
$(\varepsilon,\lambda_0=1.0005,1.3)$. Increasing $\lambda_0$, we move to locations `b'
$(\varepsilon,\lambda_0=1.0005,1.41)$,
`c' $(\varepsilon,\lambda_0=1.0005,1.425)$ and `d' $(\varepsilon,\lambda_0=1.0005,1.52)$.
And then for higher $\varepsilon$,
solution of panel `e' represents solution corresponding to location `e' $(\varepsilon,\lambda_0=1.003,1.55)$,
and panel `f' corresponds to solution for $(\varepsilon,\lambda_0=0.985,1.6)$.
The global transonic accretion solutions in the inset are represented by solid curve (online black), and solutions
which represent wind-type solutions are represented by the dotted curve (online red). The dashed
curve represents the transonic solution through which matter may pass. 
Accretion flows with any value of 
$\varepsilon$ and $\lambda_0$ in the bounded region ABDEA has three sonic points, and have a combination
of closed and global solutions (\eg Figs. 1b --- e). In the domain of multiple sonic points,
if the steady shock solution cannot be obtained, and in addition,
the solution through $x_{\rm co}$ do not connect the horizon and infinity, then
the solutions through $x_{\rm co}$ cannot be determined exactly, and so, only solutions through inner
sonic point are shown for Fig. 1d. The rough location of $x_{\rm co}$ can be ascertained
for Fig. 1d, at the location where global solution through $x_{\rm ci}$
(solid and online black), shows a maximum. An ADAF-type solution, \ie monotonic $M$ distribution through $r_{\rm ci}$,
is shown in Fig. 1e.
On the other hand,
flows with $\varepsilon$ and $\lambda_0$ in the region BCDB have two critical points, and produces only
closed topologies and therefore no global transonic solution (Fig. 1f).
Within ABDEA, the bounded region FGHF produces steady, non-dissipative shocked solutions (\eg Fig. 1c),
which are obtained from equations (\ref{mfs.eq})-(\ref{enf.eq}). The region outside BAEI and $\varepsilon \ge 1$,
there is only one sonic point (\eg Figs. 1a \& f). 
Fig. 1a, has low angular momentum and therefore produces a Bondi-type solution characterized by
a single sonic point far away from the horizon, even in presence of dissipation.
Fig. 1e, on the other hand produces a solution which is mostly subsonic and becomes transonic close to the 
horizon, and is similar to ADAF-type solutions.
Regions outside D$^{\prime}$BCD and $\varepsilon<1$  and right of the curve DI, there
exists no critical point, and consequently steady state black hole accretion is not allowed for such inner boundary
conditions. The coordinates of the important
points which represents the multiple sonic point domain in the energy-angular momentum parameter space
are A ($1.0055, 1.311$), B ($1.0, 1.385$),
C ($0.959, 1.682$), D ($1.0, 1.822$) and E ($1.0003, 1.824$).

\begin{figure}
 \begin{center}
  \includegraphics[width=12.cm]{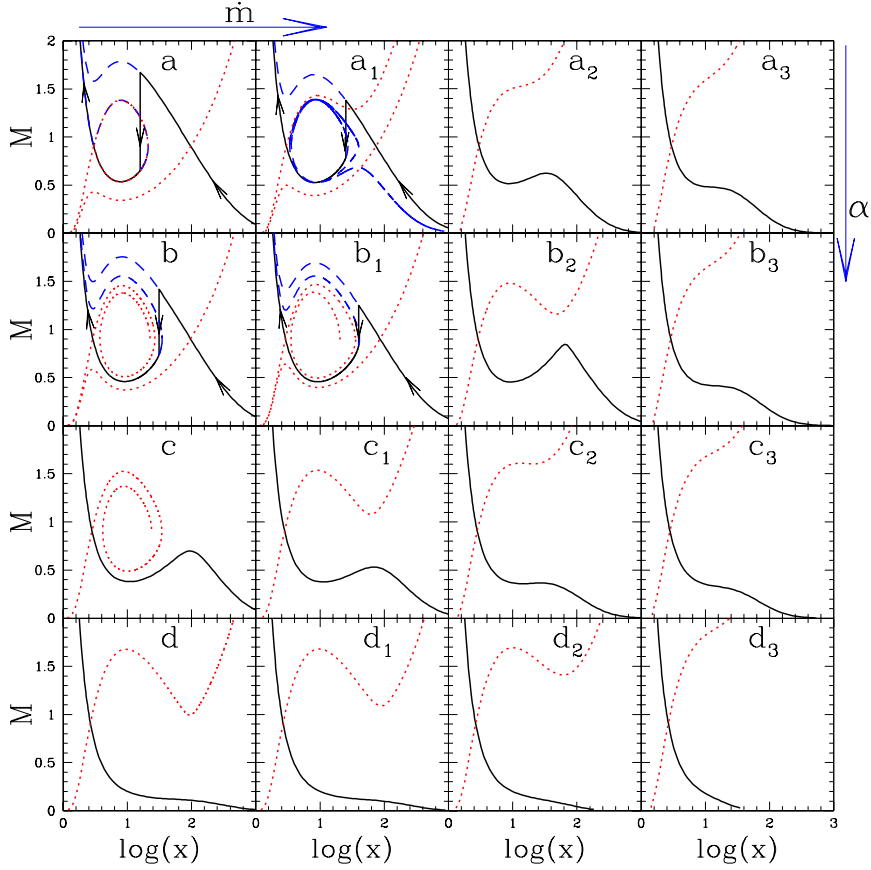}
\caption{Effect of accretion rate and viscosity parameter for boundary conditions which support steady shock
in inviscid flow.
$\varepsilon=1.001,~ \lambda_0=1.55$, $\beta=0.1$ and $\xi=1.0$ are same for all solutions.
(a) to (d) $\chi=0.0$ and $\alpha=0.0,~0.01,~0.02$ and $0.043$, respectively.
(a$_1$) to (a$_3$) $\chi=1.0$ and $\dot m=0.2,~0.8,~1.4$
(b$_1$) to (b$_3$) $\chi=1.0$ and $\dot m=0.03,~0.18,~0.9$.
(c$_1$) to (c$_3$) $\chi=1.0$ and $\dot m=0.1,~0.5,~1.0$.
(d$_1$) to (d$_3$) $\chi=1.0$ and $\dot m=0.01,~0.1,~1.0$.
Wind-type solutions are plotted as dotted curve (red online), and dashed (blue) part of the transonic solution
which is not followed by the flow in presence of shock.}
\end{center}
\label{lab:fig3}
\end{figure}

Having shown how solution of electron-proton flow depends on $\varepsilon$ and $\lambda_0$ for
given values of $\alpha$, $\beta$ and ${\dot m}$, we now present dependence of advective solutions on
$\alpha$ and ${\dot m}$ for given values of $\varepsilon$ and $\lambda_0$.
We choose flow parameters $(\varepsilon,~\lambda_0)=(1.001,1.46)$ which
produces shock-free solution with a single
$x_{\rm co}$-type sonic point in the inviscid limit, \ie for $\alpha=\chi=0$ (Fig. 2a). Panels on the left,
Figs. 2(a)---(e), represent solutions without cooling \ie $\chi=0.0$, but increasing viscosity
$\alpha=0.025$ (Fig. 2b), $\alpha=0.03$ (Fig. 2c), $\alpha=0.048$ (Fig. 2d), $\alpha=0.065$ (Fig. 2e).
We show that a shock-free solution through a single $x_{\rm co}$
in the inviscid limit (Fig. 2a), enters multi-critical point domain (Fig. 2b), and eventually
generates shock (Fig. 2c) as $\alpha$ is increased for the
same inner boundary condition \ie $\varepsilon-\lambda_0$. Further increase of $\alpha$ removes steady shock
while still being in the multi-critical point domain 
(Fig. 2d), and eventually produces a monotonic shock-free solution through $x_{\rm ci}$ or ADAF type solution.
In Figs. 2(a)-(a$_3$), the solutions are inviscid, but ${\dot m}$ is increased step by step to values
$3.5$ (Fig. 2a$_1$), $4.0$ (Fig. 2a$_2$), and $5.0$ (Fig. 2a$_3$).
For Figs. 2 (b$_1$)-(b$_3$), $\alpha=0.025$ (same as Fig. 2b), ${\dot m}$ varies from $0.7,~1.4$ and $2.1$, respectively.
For Figs. 2(c$_1$)-(c$_3$), $\alpha$ is same as Fig. 2c, but ${\dot m}$ varies from $0.2,~0.8$ and $1.6$, respectively.
The accretion rate ${\dot m}$ increases from $0.1$ in Fig. 2d$_1$, to $0.5$ in Fig. 2d$_2$ and then up to
$1.0$ in Fig. 2d$_3$, and has the same $\alpha$ as Fig. 2d.
On the other hand, $\alpha$ of Fig. 2e is used for Figs. 2(e$_1$)-(2e$_3$), but accretion rates are $\dot m=0.01$ (Fig. 2e$_1$),
${\dot m}=0.1$ (Fig. 2e$_2$) and ${\dot m}=1.0$ (Fig. 2e$_3$). Therefore, it is clear that the
very inner boundary condition
(read $\varepsilon~-~\lambda_0$) which produces a shock-free, monotonic solution with only one $x_{\rm co}$ sonic point
in the inviscid limit,
for different $\alpha$ and ${\dot m}$ will produce such varied $\lambda$ and $\Theta$ distributions that would generate
solutions comprising multiple-sonic points, shocks, or monotonic ADAF-type solutions. 
In the next figure we consider a different inner boundary condition. In Fig. 3a, we 
consider the parameters $(\varepsilon,~\lambda_0=1.001,~1.55)$, which in the inviscid limit produces shock at
$x_{\rm s}=15.936$.
The composition of the flow is $\xi=1.0$.
We increase $\alpha$ as we go vertically down, which are $\alpha=0.01$ (Fig. 3b), $\alpha=0.02$ (Fig. 3c) and $\alpha=0.043$
(Fig. 3d), but keep $\chi=0$. We turn on $\chi$, and increase $\dot m$ to the right, while keeping $\alpha$
in each row the same. In Figs. 3a$_1~\rightarrow$ 3a$_3$, $\alpha=0,~\beta=0.1 ~\chi=1$, but ${\dot m}=0.2,~0.8,~1.4$,
respectively.
In Figs. 3 (b$_1$) to (b$_3$), $\alpha=0.01, ~\chi=1$, but ${\dot m}=0.03,~0.18,~0.9$, respectively.
For Figs. 3 (c$_1$) to (c$_3$) the parameters are $\alpha=0.02, ~\chi=1$ and
${\dot m}=0.1,~0.5,~1.0$, respectively. Shocked solutions seems to be maintained for a range of $\alpha$ and
${\dot m}$, but playing with various flow parameters generates all possible solutions, including ADAF type solutions.
Figs. 3 (a)-(d$_3$) show that the closed topology through $x_{\rm ci}$ in the inviscid limit,
opens up with the increase of viscosity and cooling.

\begin{figure}
 \begin{center}
  \includegraphics[width=12.cm]{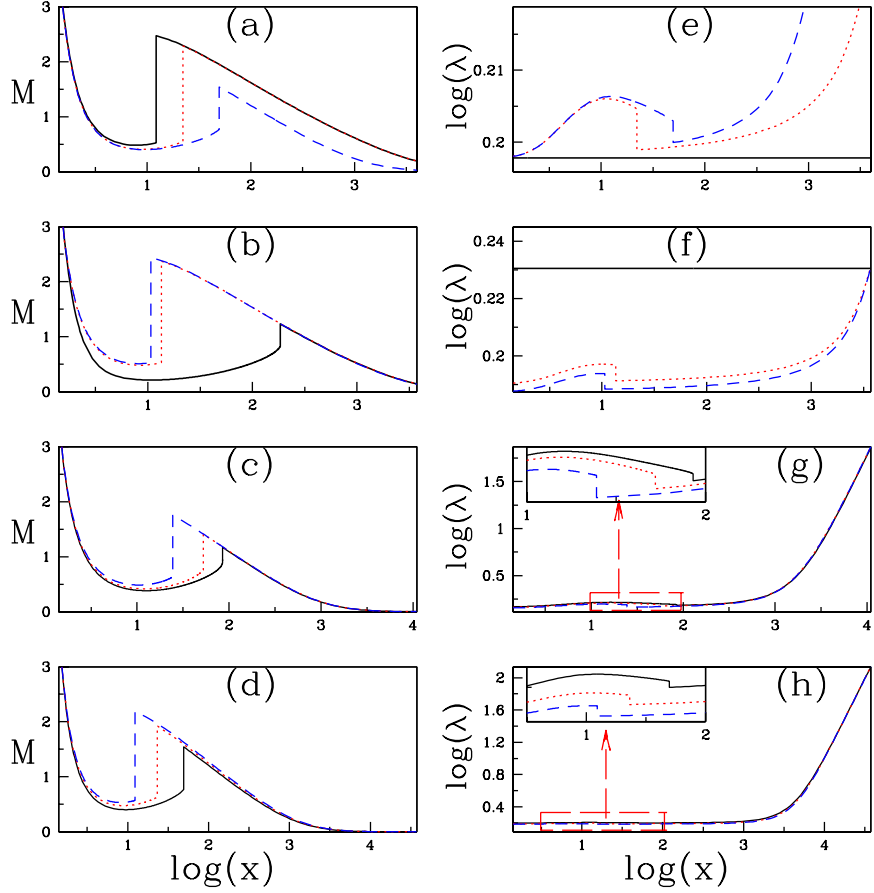}
  \caption{Variation of $M$ (a-d) and corresponding log$(\lambda)$
(e-h) with log$(x)$, for $\ep$ flow.
Solutions (a and e) are plotted for same
$\varepsilon=1.0001,~\lambda_0=1.577$.
Each curve corresponds to $\alpha=0.0,~\chi=0.0$ and a shock at $x_{\rm s}=12.2236$ (solid, online black),
$\alpha=0.01,~\chi=0.0$ and a shock at $x_{\rm s}=22.1412$ (dotted, online red)
and $\alpha=0.01,~\chi=1.0$ and a shock at $x_{\rm s}=49.4872$ (dashed, online blue).
Each curve in (b and f) is generated for $\alpha=0.0,~\chi=0.0$ with $x_{\rm s}=183.9687$ (solid, online black), 
$\alpha=0.01,~\chi=0.0$ with $x_{\rm s}=13.5259$ (dotted, online red) and $\alpha=0.01,~\chi=1.0$ with $x_{\rm s}=10.6728$
(dashed online blue), but for same outer boundary conditions
  $\lambda_{\rm inj}=1.7, ~\Theta_{\rm inj}=9.811\times10^{-2} ~\mbox{and}~\vel_{\rm inj}=1.928\times 10^{-3}$, at
  $x_{\rm inj}=3686.02$.
Plots a and e, b and f have common parameters, $\beta=0.1~\mbox{and}~\dot{m}=0.1$.
Each curve in (c and g) is plotted with $\alpha=0.0494$ and produces a shock at $x_{\rm s}=85.1545$
(solid, online black), 
$\alpha=0.0534$ and $x_{\rm s}=52.3582$ 
(dotted, online red), $\alpha=0.0545$ and $x_{\rm s}=24.5147$ (dashed, online blue) and $\chi=1.0$
but keeping other parameters,
  $\lambda_{\rm inj}=\lambda_K(x_{\rm inj})=74.12,
\Theta_{\rm inj}=0.3999~\mbox{and}~\vel_{\rm inj}=4.3568\times 10^{-5}$ fixed 
  at the outer boundary $x_{\rm inj}=10986.38$ with $\beta=0.01,~\dot{m}=0.1$.
Each curve in (d) and (h) are plotted with different accretion rates $\dot{m}=0.1$ (solid, online black), $0.3$
  (dotted, online red) and $0.5$ (dashed, online blue) and $\alpha=0.01, \beta=0.1$ are fixed.
We keep $\lambda_{\rm inj}=\lambda_K(x_{\rm inj})=136.67, \Theta_{\rm inj}=3.554~\mbox{and}~
\vel_{\rm inj}=1.1571\times 10^{-5}$ 
  are fixed at the outer boundary $x_{\rm inj}=37354.32$.
The shocks are at $x_{\rm s}=49.4872$, $23.13107$ and $12.3237$, respectively.
  }
 \end{center}
\label{lab:fig4}
\end{figure}

In Figs. 4 (a)-(h), we compare accretion solutions by varying $\alpha$, $\chi$
or ${\dot m}$ but for
either same inner boundary condition
or same outer boundary condition. In Figs. 4 (a)-(d) we plot $M$,
and in Figs. 4 (e)-(h) we plot log$(\lambda)$ with log$(x)$. Each pair of
horizontal panels show the Mach number and angular momentum distribution
of flows with same boundary condition. For example, Figs. 4 (a) and (e)
presents $M$ and log$(\lambda)$ distribution of accretion flows with
same inner boundary condition, \ie $\varepsilon=1.0001,~{\rm and}~\lambda_0=1.577$,
where each curve represents $\alpha=0.0,~\chi=0.0$ (solid, online black), $\alpha=0.01,~\chi=0.0$ (dotted, online red)
and $\alpha=0.01,~\chi=1.0$ (dashed, online blue). Clearly,  $\lambda=\lambda_0$ is a constant of motion
for inviscid and adiabatic solution.
Evidently, the shock recedes as viscosity is turned
on ($x_{\rm s}=12.2236~\rightarrow~22.1412$), and then further recedes to $x_{\rm s}=49.4872$ as the cooling
is turned on over and above the viscous dissipation. Since the effect of
viscosity is to reduce angular momentum inwards, and the effect of cooling is to reduce temperature inwards,
therefore, keeping same inner boundary and increasing viscosity and cooling implies both angular momentum and
temperature increases outwards.
Higher temperature and angular momentum means the shock front is shifted outwards. 
In Figs. 4 (b) and (f), we again compare inviscid flow (solid, online black), with flow in presence of viscosity,
\ie $\alpha=0.01,~{\rm and}~\chi=0.0$ (dotted, online red), and viscous flow in presence of
cooling ,\ie $\alpha=0.01,~\chi=1.0$ (dashed, online blue), but now the flows are launched with the same outer boundary
condition,
\ie $\lambda_{\rm inj}=1.7$, $\Theta_{\rm inj}=9.811\times10^{-2}$
and $\vel_{\rm inj}=1.928\times10^{-3}$, at the injection radius $x_{\rm inj}=3686.02$.
The accretion rate is ${\dot m}=0.1$ and $\beta=0.1$ for the flow with $\chi=1$. Since in this case we start with the same
temperature, angular momentum and velocity at the outer boundary, viscosity and cooling processes
decrease both $\lambda$ and $\Theta$
inwards. Therefore, for flows starting with same outer boundary condition, the net effect of
increasing viscosity and cooling is to reduce both the centrifugal force and the pressure,
so the shock front shifts closer to the horizon for viscous fluid with and without cooling, compared to the
inviscid flow. 
In Figs. 4c \& 4g, we compare viscous flows in presence of cooling ($\beta=0.01,~\dot{m}=0.1~\mbox{and}~\chi=1.$),
and starting with the same outer boundary condition   
($\lambda_{\rm inj}=\lambda_{\rm K}(x_{\rm inj})=74.12,
\Theta_{\rm inj}=0.3999~\mbox{and}~\vel_{\rm inj}=4.3568\times 10^{-5}$ at $x_{\rm inj}=10986.38$),
but now for different viscosity parameters, namely, $\alpha=0.0494~(\mbox{solid online
black}),~0.0534~ (\mbox{dotted, online red}),~0.0545~(\mbox{dashed, online blue})$. In the previous panel,
the solution started with sub-Keplerian flow at the outer boundary. In this figure we compare flows with different
$\alpha$, but in presence of same cooling parameters, and starting with Keplerian angular momentum (indicated by suffix
K)
at the outer boundary.
Increase of $\alpha$ shows that the
reduction of $\lambda$ causes the shock to shift inward (see inset), even in presence of cooling.
This shows that $x_{\rm s}$ reduces with increasing $\alpha$ for flows starting with same outer boundary condition.
In the next pair of panels Figs. 4 (d) and (h), we compare accretion solutions starting with the same outer boundary
condition ($\lambda_{\rm inj}=\lambda_{\rm K}(x_{\rm inj})=136.67, \Theta_{\rm inj}=3.554~\mbox{and}~
\vel_{\rm inj}=1.1571\times 10^{-5}$ at $x_{\rm inj}=37354.32$), and same $\alpha=0.01$, but different ${\dot m}=0.1$
(solid, online black), ${\dot m}=0.3$
(dotted, online red) and ${\dot m}=0.5$ (dashed, online blue),
in other words, we study the effect of cooling in a viscous flow
starting with the same outer boundary condition.
In this case, although cooling do not directly affect the angular momentum equation (\ref{dlmd.eq}), but
it affects the entropy equation and therefore the thermal energy, which reduces the post-shock
pressure. As a result shock moves inwards with the increase of cooling. Since kinematic
viscosity parameter (\ie $\nu$, defined in equation \ref{angm.eq}) depends on both $\alpha$ and $a^2$,
so cooling processes will affect $a$ and thereby in an indirect
way cooling processes will affect the angular distribution too, as is shown in Fig. 4h. 

\begin{figure}
 \begin{center}
  \includegraphics[width=10.cm]{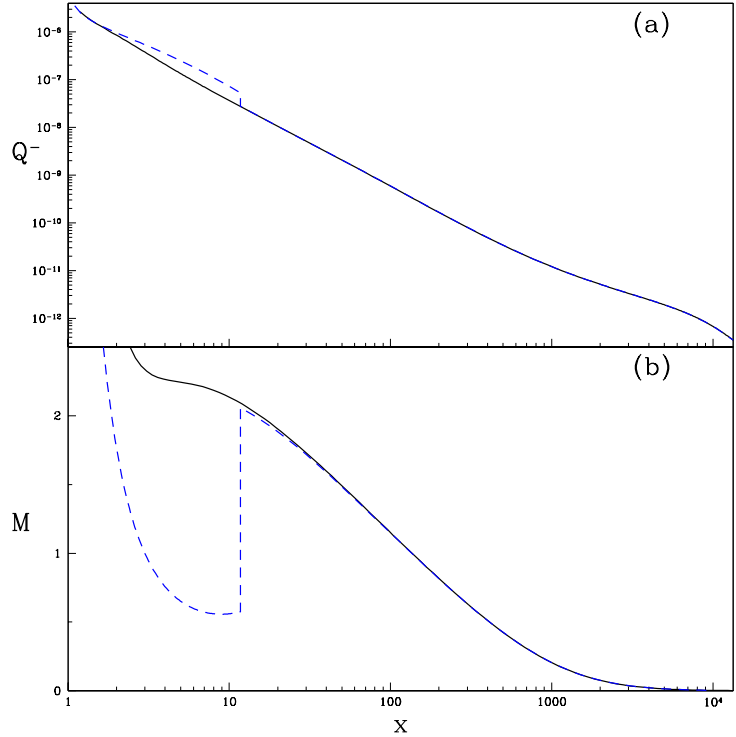}
  \caption{Variation of emissivity per unit mass $Q^-$ (a) and $M$ (b) with
$x$. Each curve represents a shock free (solid, online black) and shocked accretion solution
(dashed, online blue) generated with the same
$\lambda_{\rm inj}=\lambda_K(x_{\rm inj})=82.51$ at $x_{\rm inj}=13614.44$,
$\varepsilon=1.0005,~\beta=0.01~\mbox{and}~\dot{m}=0.1$. The shocked solution is generated with $\alpha=0.05$
and the shock free for $\alpha=0.0505$. Both the solutions are for $\ep$ flow.
Luminosities for shock free and shocked solutions are $\ell=1.40\times10^{-4}~\mbox{and}~1.67\times 10^{-4}$,
respectively. 
}
 \end{center}
\label{lab:fig5}
\end{figure}

We compare the emissivity per unit mass \ie $Q^-$ (Fig. 5a) and the Mach number $M$ (Fig. 5b)
between a shock free (solid, online black) and a shocked (dashed, online blue) accretion
solution, starting with the same outer boundary condition $\varepsilon=1.0005, ~\beta=0.01~\mbox{and}~\dot{m}=0.1$ and
$\lambda_{\rm inj}=\lambda_{\rm K}(x_{\rm inj})=82.51$ at $x_{\rm inj}=13614.44$.
The shocked solution is generated with $\alpha=0.05$
and the shock-free for $\alpha=0.0505$. Although the radiative output is almost similar
in the outer regions, but the post shock flow is more luminous. The over all luminosity of shocked solution is more compared
to the shock free solution, even for flow
with same composition, outer boundary condition and ${\dot m}$. Although we can choose to compare even hotter shock free
and shocked solutions, but the greater Comptonization efficiency of shocked solution will in general be more luminous and will
also produce a harder spectrum.

\begin{figure}
 \begin{center}
  \includegraphics[width=12.cm]{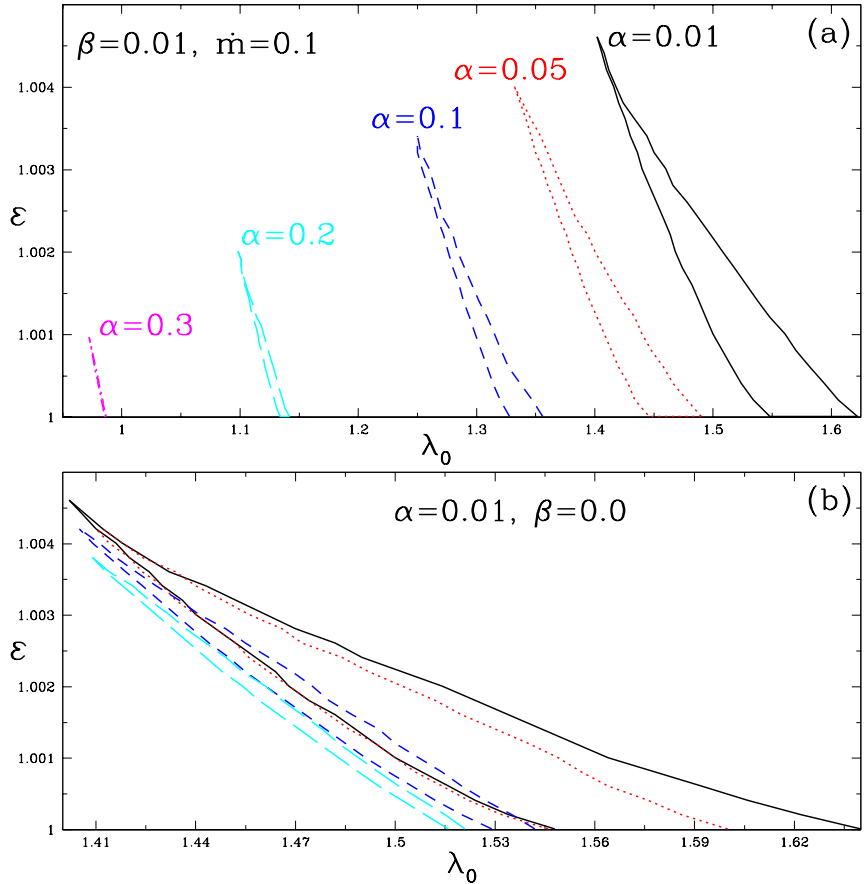}
  \caption{(a) Non-dissipative shock parameter space with general Bernoulli parameter ($\varepsilon$) versus specific
  angular momentum ($\lambda_0$) for $\alpha=0.01~(\mbox{solid, online black}),~0.05~(\mbox{dotted, online red}),~
  0.1~(\mbox{dashed, online blue})$, $0.2~(\mbox{long-dashed, online cyan})~\mbox{and}~0.3~
(\mbox{dotted-dashed, online magenta})$ 
 and with constant cooling parameters, $\dot{m}=0.1~\mbox{and}~ \beta=0.01$. (b)
$\varepsilon$-$\lambda_0$ parameter space for non-dissipative shock for different 
$\dot{m}= 0.1~(\mbox{solid, online black}),~1.0~(\mbox{dotted, online red}),~10.0~(\mbox{dashed, online blue})~
\mbox{and}~20.0~(\mbox{long-dashed, online cyan})$ and keeping $\alpha=0.01~\mbox{and}~\beta=0.0$ fixed. Both the plots are
for $\ep$ flow.}
 \end{center}
\label{lab:fig6}
\end{figure}

\begin{figure}
 \begin{center}
  \includegraphics[width=12.cm]{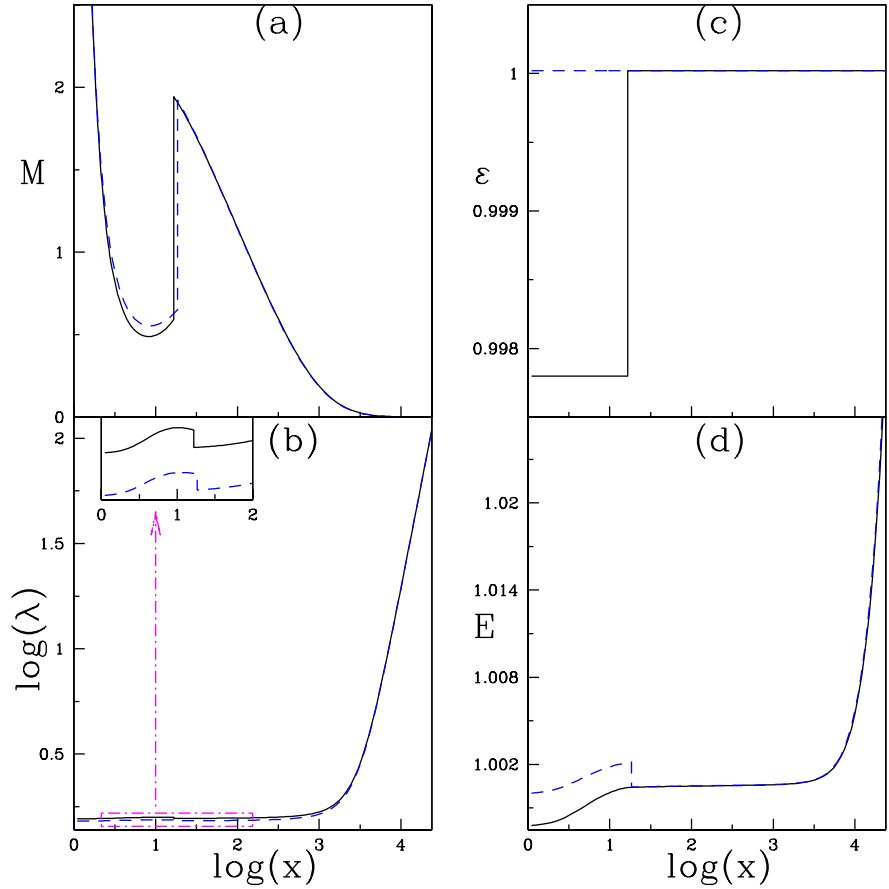}
\caption{Variation of various physical quantities of the accretion flow such as Mach number $M$ in plot (a),
specific angular momentum $\lambda$ 
in plot (b), general Bernoulli parameter $\varepsilon$ in plot (c), and grand specific energy $E$ in plot (d)
are plotted with radial distance $log(x)$.
In all plots solid (online black) curves represent dissipative shock and dashed (online blue) curves represent non-dissipative
shock solutions and having shock locations at $16.61~\mbox{and}~18.51$, respectively. The solutions
are generated for $\alpha=0.01$, $\beta=0.01$, and ${\dot m}=1$.
Outer boundary condition is $\varepsilon=1.000021, \lambda_{\rm inj}=\lambda_K(x_{\rm inj})=109.19$ at 
$x_{\rm inj}=23842.73$. For dissipative shock, the energy 
dissipated at the shock is $\Delta\varepsilon=0.00221$.}
\end{center}
\label{lab:fig7}
\end{figure}

In Fig. 6a, we plot the non-dissipative shock parameter space $\varepsilon-\lambda_0$ of accretion flow in presence of
cooling ($\chi=1,~\beta=0.01,~{\rm and}~{\dot m}=0.1$), but for different viscosity parameters
$\alpha=0.01~(\mbox{solid, online black}),
~0.05~(\mbox{dotted, online red}),~
0.1~(\mbox{dashed, online blue})$, $0.2~(\mbox{long-dashed, online cyan}),~\mbox{and}~0.3~
(\mbox{dotted-dashed, online magenta})$. One can obtain steady shocks at $\alpha>0.3$ in presence of cooling too.
In Fig. 6b, we plot shock parameter space $\varepsilon-\lambda_0$ of accretion flow for a particular $\alpha=0.01$
but for different accretion rate ${\dot m}=0.1~(\mbox{solid, online black}),~1.0~(\mbox{dotted, online red}),
~10.0~(\mbox{dashed, online blue})$ and $~20.0~(\mbox{long-dashed, online cyan})$, however, with synchrotron
processes ignored, \ie $\beta=0$. It is interesting to note that shocked accretion solution can be obtained
for fairly high $\alpha$ and ${\dot m}$. 

\subsubsection{Dissipative shock in accretion flow}
\begin{figure}
 \begin{center}
  \includegraphics[width=12.cm]{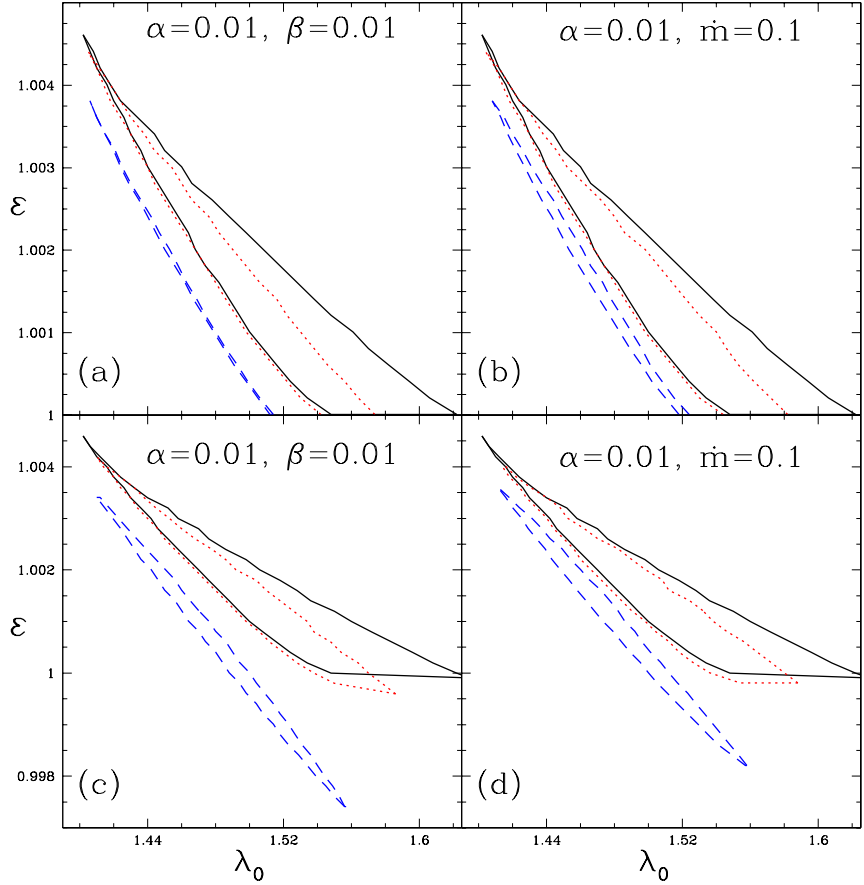}
  \caption{$\varepsilon-\lambda_0$ parameter space for non-dissipative (a and b) and 
  dissipative (c and d) shocks in the flow. Bounded regions in (a, c) are plotted
with different accretion rates, $\dot{m}=0.1
  ~(\mbox{solid line}),~1.0~(\mbox{dotted line})~\mbox{and}~10.0~(\mbox{dashed line})$ but keeping 
  $\beta=0.01$ fixed, and regions in (b, d) are plotted with different $\beta=0.01~(\mbox{solid line}),~0.1
  ~(\mbox{dotted line})~\mbox{and}~1.0~(\mbox{dashed line})$ but keeping $\dot{m}=0.1$ fixed. 
  All plots are for the same viscosity parameter, $\alpha=0.01$. 
  }
 \end{center}
\label{lab:fig8}
\end{figure}

All shocked solutions presented in the preceding subsection have been examples of non-dissipative shocks
obtained by solving equation (\ref{suq.eq}). However, in section 2.1.2, we have discussed if the energy flux
is not conserved and still there is a shock, then that would be considered as a dissipative shock.
The conditions for a dissipative shock is presented by equation (\ref{iso.eq}). In Figs. 7a-7d,
we compare accretion solutions starting with the same outer boundary condition
[$\varepsilon=1.000021, \lambda_{\rm inj}=\lambda_K(x_{\rm inj})=109.19$ at $x_{\rm inj}=23842.73$], same
$\alpha=0.01$, $\beta=0.01$ and ${\dot m}=1$, but one
solution harbours dissipative shock (solid online black) and the other solution harbours
non-dissipative shock (dashed online blue). We compare $M$ (Fig. 7a), $log(\lambda)$ (Fig. 7b),
${\varepsilon}$ (Fig. 7c)
and $E$ (Fig. 7d) as a function of $log(x)$.
Conservation of $E$ across the shock is equivalent to a discontinuous decrement in $\varepsilon$ across the shock
(solid curve in Fig. 7c),
corresponding to an energy dissipation of $\Delta \varepsilon=0.00221$. Compared to the non-dissipative shock,
the dissipative shock forms closer to the horizon because of $\Delta \varepsilon$ released at the dissipative shock.
It is also clear from Figs. 7c and 7d,
that $\varepsilon$ is a constant of motion in presence of
viscous dissipation and cooling processes, and $E$ is not.
In Figs. 8a-8d, we plot the $\varepsilon - \lambda_0$ shock parameter space for non-dissipative shock (Figs. 8a and 8b)
and dissipative shocks (Figs. 8c \& 8d). In Figs. 8a and 8c, the steady shock parameter space
are bounded regions for ${\dot m}=0.1$ (solid, online black), ${\dot m}=1$ (dotted, online red),
and ${\dot m}=10$ (dashed, online blue), all the plots are generated for given values
of $\alpha=0.01$ and $\beta=0.01$. In Figs. 8b and 8d, the shock parameter space are the bounded
regions characterized by $\beta=0.01~(\mbox{solid line}),~0.1
  ~(\mbox{dotted line})~\mbox{and}~1.0~(\mbox{dashed line})$ but keeping $\alpha=0.01,\mbox{ and }\dot{m}=0.1$ fixed.
So in Figs. 8a and 8c, we compare non-dissipative and dissipative shocks for same proportion of synchrotron losses but different
mass supply, 
and in  Figs 8b and 8d, we compare non-dissipative and dissipative shocks for same mass supply but different synchrotron losses.
We kept the viscosity parameter same to see the effect of ${\dot m}$ and $\beta$. Clearly the parameter ranges for dissipative
steady shock is larger.
Evidently,
the combined parameter space for both non-dissipative and dissipative shocks is quite significant. 
From Figs. 6a-6b and 8a-8d, it is clear that steady shock may exist for fairly extreme flow parameters
like super Eddington accretions rates, $\alpha \gsim 0.3$, and fairly high magnetic energy. 

 \subsubsection{Hyper-accretion rate} 
\begin{figure}
 \begin{center}
  \includegraphics[width=12.cm]{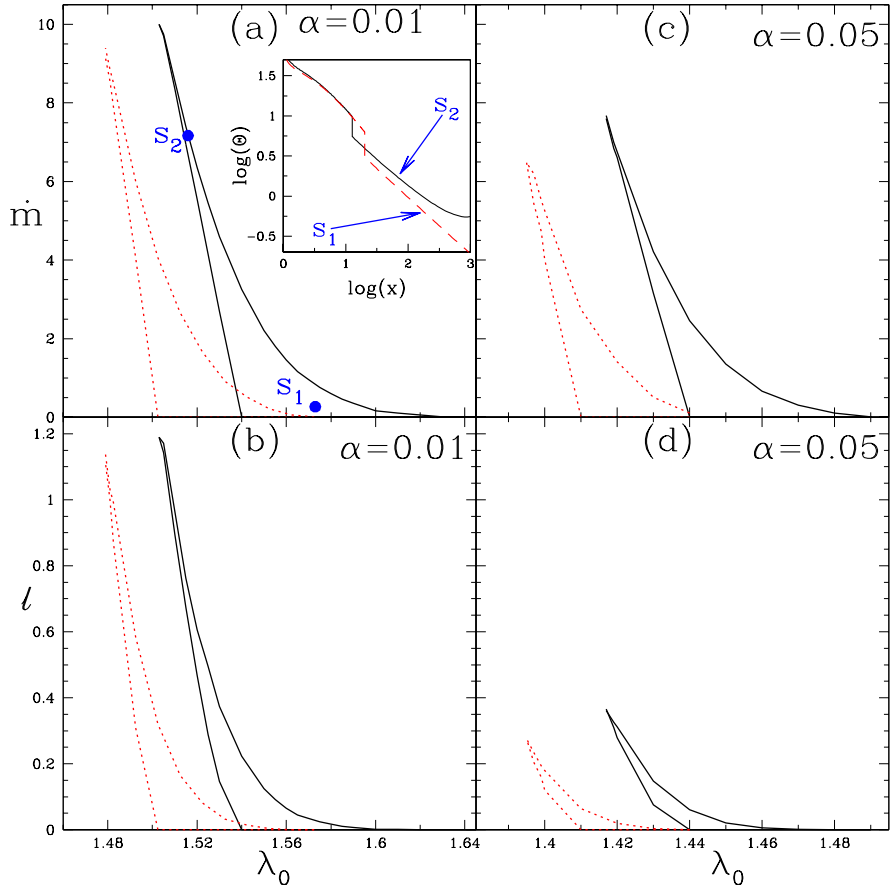}
  \caption{Shock parameter space $\dot{m}-\lambda_0$ in panels (a and c) and corresponding 
luminosities $\ell$ in panels (b and d). Plots (a and b) are generated for  viscosity parameter,
$\alpha=0.01$ and plots (c and d) are generated for $\alpha=0.05$. Each curve corresponds to,
$\varepsilon=1.0001~(\mbox{solid line})~\mbox{and}~1.001~(\mbox{dotted line})$. We keep
$\beta=0.01$ same for all the plots. Inset in (a),
$\Theta$ as a function of $x$in log-log scale. Solution S$_1$ (shown in ${\dot m}-\lambda_0$ space)
$\varepsilon=1.0001$, $\lambda_0=1.57$, $\alpha=0.01$, $\beta=0.01$, ${\dot m}=0.1$,
$x_{\rm s}=20.02547$ and disc luminosity $\ell=2.766 \times 10^{-4}$. 
Solution S$_2$ (shown in ${\dot m}-\lambda_0$ space) is for 
$\varepsilon=1.0001$, $\lambda_0=1.515$, $\alpha=0.01$, $\beta=0.01$, ${\dot m}=7.0$,
$x_{\rm s}=12.73524$ and disc luminosity $\ell=0.677$. All the plots are for $\ep$ flow. 
}
 \end{center}
\label{lab:fig9}
\end{figure}

In Figs. 9a and 9c, we plot shock parameter spaces \ie bounded region of $\dot{m}-\lambda_0$ and
the corresponding $\ell-\lambda_0$ for a viscosity parameter $\alpha=0.01$,
and in Figs. 9b and 9d, the $\dot{m}-\lambda_0$ and $\ell-\lambda_0$ shock parameter spaces for $\alpha=0.05$.
Each curve are plotted for $\varepsilon=1.0001$ (solid, online black) and $\varepsilon=1.001$ (dotted, online red).
For all the plots $\chi=1$ and $\beta=0.01$. It is interesting to note that a steady shock can form in accretion
flow even for super Eddington accretion rate, and can also radiate on or above Eddington luminosity.
The efficiency of conversion of accretion power to radiation also varies, for example we consider two accretion
solutions corresponding to $\varepsilon=1.0001$ and $\alpha=0.01$,
and $\lambda_0=1.57$ (S$_1$) and $\lambda_0=1.515$ (S$_2$). The dimensionless temperature
$\Theta$ of the two solutions are plotted with $x$ in log-log scale and are presented in the inset of Fig. 9a.
S$_1$ corresponds to ${\dot m}=0.1$ and $\ell=2.766\times10^{-4}$ and S$_2$ corresponds
to ${\dot m}=7$ and $\ell=0.677$. The radiative efficiency defined as $\ell/{\dot m}$ of S$_1$
is $\lsim 10^{-3}$, while for S$_2$
the efficiency is $\sim 0.1$. Therefore, the range of radiative efficiency obtained from our solutions, spans
from radiatively inefficient advective flow to radiatively luminous regime, and solely depends on the outer
boundary conditions. So the cycle of low luminosity to luminous but intermediate hard states in microquasars can
be addressed if all the solutions in the advective regime be considered.

\subsection{Effect of composition}

\begin{figure}
 \begin{center}
  \includegraphics[width=12.cm]{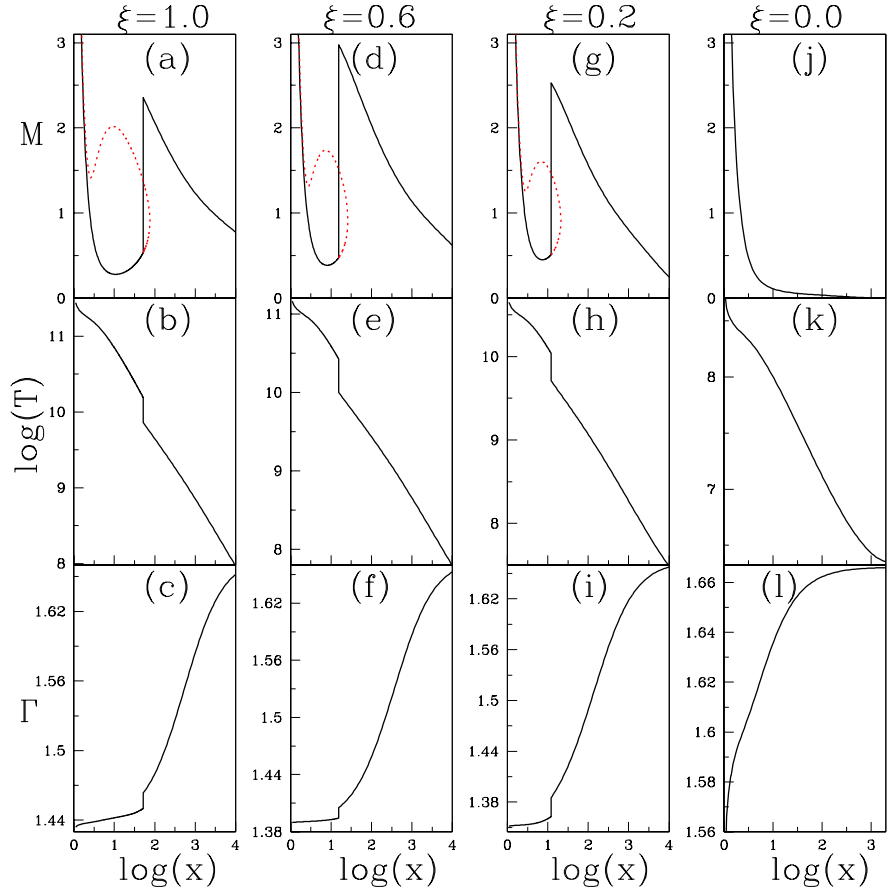}
  \caption{Variation of Mach number $M$ (a, d, g, j), $log(T )$ (b, e, h, k) and adiabatic index $\Gamma$
(c, f, i, l) with radial distance $log(x)$.
These solutions are plotted with disc parameters, $\varepsilon = 1.000001$, $λ_0 = 1.63$, $\alpha = 0.01$, $\beta = 0.01$,
${\dot m} = 0.001$ and
$\xi = 1.0$ (a-c), $0.6$ (d-f), $0.2$ (g-i), $0.0$ (j-l). The plots with $\xi = 1.0,~ 0.6,~ {\rm and}~ 0.2$
having shock locations at $x_{\rm s} = 51.82,~ 15.43,~ 12.06$, respectively.
For $0\leq \xi \leq 0.157$, shock solution does not exist for these parameters.}
 \end{center}
\label{lab:fig10}
\end{figure}

\begin{figure}
 \begin{center}
  \includegraphics[width=12.cm]{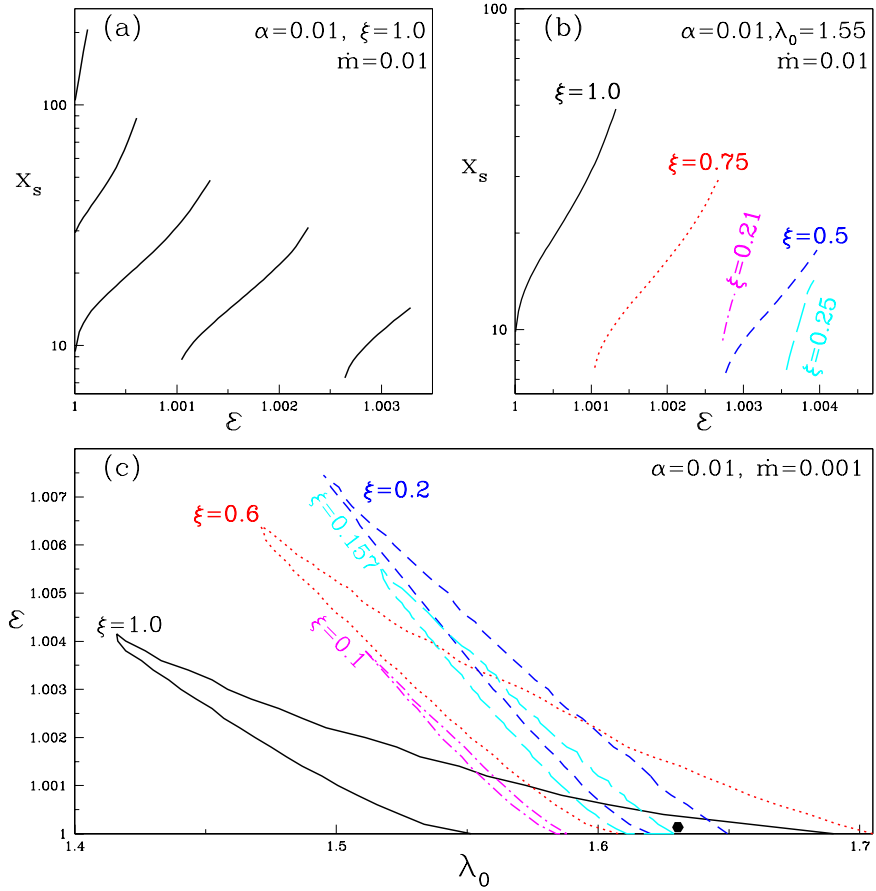}
\caption{(a) Variation of $log(x_{\rm s})$ with $\varepsilon$. From left to right, \ie $\lambda_0=1.65$ to $1.45$ 
with decrement $d\lambda_0=0.05$, with parameters, $\alpha=0.01, ~\xi=1.0 ~\beta=0.01~~\dot{m}=0.01,
\mbox{ and} ~\chi=1$.
(b) Variation of $log(x_{\rm s})$ with $\varepsilon$. The composition of the flow are $\xi=1.0$ (solid, online black),
$\xi=0.75$ (dotted, online red), $\xi=0.5$ (dashed, online blue),
$\xi=0.25$ (long dashed, online cyan), $\xi=0.21$ (dashed-dotted, online magenta). Other parameters are
$\alpha=0.01$,~ $\lambda_0=1.55$,~ $\beta=0.01~\mbox{and}~\dot{m}=0.01$.
(c) $\varepsilon-\lambda_0$ parameter space for steady shocks for $\alpha=0.01$, $\beta=0.01$ and
${\dot m}=0.001$. Each curve corresponds to $\xi=1.0$ (solid, online black), $\xi=0.6$ (dotted, online red),
$\xi=0.2$ (dashed, online blue), $\xi=0.157$ (long dashed, online cyan), $\xi=0.1$ (dashed-dotted,
online magenta). The black dot near the bottom show the parameters for which solutions of Fig. 10a-10l
are plotted.
}
\end{center}
\label{lab:fig11}
\end{figure}
All previous figures were for $\ep$ flow. In Figs. 10a-10l, we compare flow variables of different
$\xi$. All the plots are generated for $\varepsilon=1.000001$, $\lambda_0=1.63$, $\alpha=0.01$,
$\beta=0.01$, and ${\dot m}=0.001$. We change the composition as we go to the right \ie $\xi=1$ (Figs. 10a-10c),
$\xi=0.6$ (Figs. 10d-10f), $\xi=0.2$ (Figs. 10g-10i) and $\xi=0.0$ (Figs. 10j-10l). And we change the flow
variable as we go vertically down \ie $M$ (Figs. 10a, 10d, 10g, 10j), $log(T)$ (Figs. 10b, 10e, 10h, 10k),
adiabatic index $\Gamma$ (Figs. 10c, 10f, 10i, 10l). These solutions
correspond to the location marked by the black dot in parameter space shown in Fig. 11c.
Once again we show that, like our previous papers \citep{c08,cr09,cc11,cmggkr12,kscc13}, the temperature of the
flow decreases with the decrease of $\xi$, but makes the flow thermally more relativistic (\eg $\Gamma_{\xi=1}
>\Gamma_{\xi=0.6}>\Gamma_{\xi=0.2}$
in Figs. 10c, 10f, 10i) because of the reduced inertia of the flow over compensates the reduced thermal energy.
However, if $\xi<0.2$, the temperature is so low (Fig. 10k) that the reduction in proportion of protons cannot compensate
and the reduced temperature of the flow, and it becomes thermally less relativistic ($\Gamma_{\xi=0}>\Gamma_{\xi\neq0}$).
Let us now
investigate how the shock location behaves with the variation of $\varepsilon$ and $\lambda_0$.
In Fig. 11a, we plot $log(x_{\rm s})$  with $\varepsilon$ for parameters $\alpha=0.01, ~\xi=1.0 ~\mbox{and} ~\chi=1$.
From left to right, curves represent $\lambda_0=1.65$ to $1.45$ 
with decrement $d\lambda_0=0.05$. This shows that for a given $\xi$, for low $\varepsilon$ steady state shock will form 
if $\lambda$ is high and vice versa. This shows that the accretion shock occurs due to presence of the centrifugal barrier
as well as high temperature of the flow. Higher $\varepsilon$ is symptomatic of hotter flow
so it can support steady shocks at lower angular momentum, while flows with lower
$\varepsilon$ needs higher angular momentum to produce shocks.
In Fig. 11b, we plot  $log(x_{\rm s})$  with $\varepsilon$ for parameters
$\alpha=0.01,~\lambda_0=1.55,~\beta=0.01~\dot{m}=0.01, ~\mbox{and} ~\chi=1$, but now each curve represent
$\xi=1.0$ (solid, online black),
$\xi=0.75$ (dotted, online red), $\xi=0.5$ (dashed, online blue),
$\xi=0.25$ (long dashed, online cyan), $\xi=0.21$ (dashed-dotted, online magenta). As $\xi$ is decreased,
the flow becomes more energetic and therefore shock forms at higher $\varepsilon$. Since the flow is thermally the
most relativistic when $\xi \sim 0.25$, so any
further decreases in $\xi$ makes the flow, as well as, the shock to be less energetic, and hence $x_{\rm s}$ shifts
towards lower $\varepsilon$. Similar to the shock location itself, the shock parameter space has a similar
tendency. In Fig. 11c, the parameters
used for all $\xi$ are $\alpha=0.01$, $\beta=0.01$ and ${\dot m}=0.001$.
The shock parameter space for $\xi=1$ (solid, online black) is in the lower $\varepsilon$,
lower $\lambda_0$ range, but reduction of protons shifts the shock parameter space to higher $\varepsilon$
and higher $\lambda_0$ range ($\xi=0.6$ dotted, online red; $\xi=0.2$ dashed, online blue). 
However, further reduction of $\xi$ makes the flow less energetic and shift towards left
($\xi=0.157$ long dashed, online cyan; $\xi=0.1$ dashed, online magenta). The black dot show the parameters
for which Figs. 10a-10k were generated. For these parameters no shock exist for $\xi<0.157$.

\section{Discussion and Concluding Remarks}

In this paper we presented the solutions of accretion flow in presence of dissipative processes like viscosity
and various cooling processes, where the accretion disc fluid is described by variable $\Gamma$ EoS,
and multiple species of particles. As far as we know, such an effort
has not been undertaken before in the context of black hole accretion.
Presently, we considered only a Schwarzschild BH. If we had used a Kerr BH,
then the length scale would get reduced. The location of the horizon shifts from
$2GM/c^2$ for non-spinning black hole to $GM/c^2$ for maximally rotating black hole
and similarly other locations like marginally stable orbit, marginally bound orbit etc. Since matter enters
deeper into the gravitational well for spinning black holes, the accreting matter
becomes much hotter than those for non-spinning or slowly spinning black holes.
The multiple critical point range of parameter space shifts to higher
energy but lower angular momentum part of the parameter space.
The counter spinning black holes, on the other hand, makes the flow even colder.
However, between maximally counter rotating flow and Kerr parameter of 0.5
co-rotating flow, the difference is not significantly large.
For maximally
co-rotating flow, the temperature near the horizon is about an order of magnitude higher than
that of slowly spinning and non-rotating blackholes. This would lead to much higher radiative
efficiency. So the maximum efficiency for non-rotating black holes
we got in this paper, was close to 10 \% of accretion
energy, but the radiative efficiency can go up to 40 \% for maximally rotating black hole.
 
In this paper we have presented the equations of motion in details, identified the
constant of motions, and the exact methodology to solve these equations.
It is well known, that black hole accretion is transonic, and the location
of sonic or critical
point and the value of $\lambda_c$ for a given boundary condition are eigen values of the problem.
We obtained the location of the sonic points as eigenvalues, by extending the Frobenius
expansion methodology of \citet{bl03}
to a flow described by variable $\Gamma$ EoS. Although we were inspired by \citet{bl03},
to find the asymptotic values of the flow variables close to the horizon, but
unlike \citet{bl03} we have not used the free fall condition
close to the horizon, in order to find the exponents of the series expansion. Rather we used the
constants of motion to find them.
Consequently, the exponents of the series expansion
obtained are different than those obtained by \citet{bl03}.
We have also presented a subsection on shocks, in order to pin point the exact form of the
conserved quantities across a thin shock. In the process, we also pointed out that
conservation of the Bernoulli parameter or the grand energy, which are constants of motion
for inviscid or viscous flow, respectively, becomes a case of dissipative shock when
both viscosity and cooling are considered.

We obtained solutions for all possible boundary conditions, and generated shock free smooth
solutions, as well as, shocked accretion solutions.
We showed that shock free smooth solutions are of two types
(i) low $\lambda$ at the outer boundary or Bondi type solutions with one outer sonic point, and (ii) high $\lambda$
at the outer boundary, or,
ADAF type solutions through inner sonic point. In Figs. 1, we presented the entire
$\varepsilon - \lambda_0$ parameter space, and all type of solutions possible.
We also showed that a Bondi type solution in low $\lambda_0$, $\alpha$, and ${\dot m}$,
may develope multiple sonic points and shocks as we change the boundary condition
or change $\alpha$, ${\dot m}$ etc. But a boundary condition which produces a transonic
solution through inner sonic point in the inviscid limit, will not develope a shock for any value
of $\alpha$, $\beta$, and ${\dot m}$ (Figs. 2 and 3).
We also showed that increasing viscosity or cooling moves the shock closer to the
horizon, provided the flow is launched from with the same outer boundary condition (Figs. 5 and 6).

We have considered bremsstrahlung and synchrotron processes as the
dominant cooling processes. The inverse Comptonization process has been taken into consideration
through a fitting function presented by \citet{kcm14}, where the fitted function is the  
Compton efficiency for all kind of seed photons (see Kumar \etal 2014, for details).
We have considered this fitted function as generic, as an obvious effort to simplify things
related to Comptonization but nonetheless to incorporate some effects of it in the solution. 
Since the post shock disc has a jump in temperature and density, it is puffed up, and can hence
intercept additional photons from the post shock disc and Comptonize it. A shock free disc
has smooth solutions and therefore will not be able to intercept additional photons, reducing the 
Comptonization efficiency. As a result we found that the shocked disc is more luminous
than the shock free disc, even when they start with the same outer boundary condition (Figs. 5a and b).
We have shown that shock solutions can be found for high enough viscosity, as well as, very high
accretion rates (Figs. 6, 8-9). We have compared the shock parameter space for both non-dissipative
shocks and dissipative shocks, and if the total shocked domain in the $\varepsilon-\lambda_0$ parameter space
is considered then it is indeed quite significant. 
Furthermore Fig. 9 shows a very interesting phenomena in which,
for high accretion rate and moderate viscosity
parameter values, luminosities of up to and over Eddington limit is possible.
This is very interesting, because Fig. 9 show that luminosities of
depending on the accretion rate
radiatively inefficient and luminous regimes both can be achieved
by tuning the matter supply at the outer boundary. Figure 9, further show that
luminosities up to
$\ell \sim 10^{39}$erg s$^{-1}$ for BHCs of $M\sim 10M_{\odot}$ \ie
stellar mass black holes,
and $\ell \sim 10^{46}$erg s$^{-1}$ for BHCs of $M\sim 10^8 M_{\odot}$
for super massive black holes can be achieved in advective and shocked accretion domain
even when only non-rotating
black holes are assumed. This means higher luminosities may be achieved for shocked accretion flow if
Kerr black holes are considered.
Furthermore, since super-Eddington
accretion is possible, then the growth of the central mass also comes
into the ambit of future study. Simple minded estimates show,
that a black hole of $10M_{\odot}$ will increase its mass by 10 \% in about
$5$ Myr, if it continues to accrete at $10 {\dot M}_{\rm Edd}$.

One may wonder what difference would it make if fixed
$\Gamma$ equation of state is used. The main difference between solutions obtained
by using a correct EoS which produces temperature dependent
adiabatic index and a fixed adiabatic index EoS (only correct for low and
ultra-relativistic temperatures see Ryu et. al. 2006) is that, the latter either under estimates or
over estimates temperature of the flow. A fixed $\Gamma$ EoS
with $\Gamma=4/3$ will over estimate the temperature of the flow, and
$\Gamma=5/3$ will under estimate it all through the flow as was
shown by \citet{rcc06}. This would invariably affect the local sound speed,
messing up with the location of sonics points. Location and strength of shocks
(if it forms) will be affected too, and thereby the radiated power.
In the inviscid limit the energy-angular
momentum parameter space was compared before \citep{c08}, and it showed
that the fixed $\Gamma$ EoS produces too high energies. In other words,
when observations would be matched with solutions, we would be predicting
wrong temperatures and densities. It is also important to take the composition
into account. We have shown that
flow with EoS of similar particles ($\equiv \xi=0$),
is physically similar to an $\el$ flow, and is a very low energetic flow with temperatures
which are orders of magnitude less
than flows composed of baryons and leptons.
In other words, a flow described by EoS of similar particle
is unlikely to be realized in nature, so for relativistic astrophysical
fluid one must study flows using EoS with {\bf $1 \geq \xi > 0$}.

\section*{Acknowledgment}
The authors acknowledge the anonymous referee for fruitful suggestions to improve the quality of the paper.

{}

\end{document}